\documentclass[11pt]{article}

\usepackage{graphicx}
\usepackage{amssymb,amsmath,amsfonts,amsthm}
\usepackage{dcolumn}
\usepackage{bm}
\usepackage[figuresright]{rotating}
\usepackage{color}
\usepackage{multirow}

\usepackage{JASA_manu}
\usepackage{natbib}

\newtheorem{lemma}{Lemma}
\newtheorem{proposition}{Proposition}
\newtheorem{corollary}{Corollary}
\newtheorem{remark}{Remark}

\begin{document}
\title{Constrained Nonlinear and Mixed Effects Differential Equation Models for Dynamic Cell Polarity Signaling}
\author{Zhen Xiao\\
Biogen Inc.\\
Cambridge, MA 02142, USA
\and
Nicolas Brunel\\
ENSIIE \& Laboratoire de Math\'ematiques et Mod\' elisation d'Evry\\
UMR CNRS 8071, Universite d'Evry, France.
\and
Zhenbiao Yang\\
Center for Plant Cell Biology, Botany and Plant Sciences Department\\
University of California, Riverside, CA 92521, USA
\and
Xinping Cui$^{*}$\\
Department of Statistics\\
University of California, Riverside, CA 92521, USA\\
\textit{email:} xinping.cui@ucr.edu
}

\maketitle

\newpage

\mbox{}
\vspace*{2in}
\begin{center}
\textbf{Author's Footnote:}
\end{center}
Zhen Xiao was a PhD student in the Department of Statistics at University of California, Riverside. He is now a senior biosatistician at Biogen Inc, Cambridge, MA, 02142 (email: nehzxiao@gmail.com); and Nicolas Brunel is an associate professor in ENSIIE \& Laboratoire de Math\'ematiques et Mod\' elisation d'Evry UMR CNRS 8071, Universite d'Evry, France (email: nicolas.brunel@ensiie.fr); and Zhenbiao Yang is a professor in the Department of Botany and Plant Sciences and the Center for Plant Cell Biology and Institute for Integrative Genome Biology at University of California, Riverside, CA 92521, USA (email: zhenbiao.yang@ucr.edu); and Xinping Cui is a professor in the Department of Statistics and the Center for Plant Cell Biology and Institute for Integrative Genome Biology at University of California, Riverside, CA 92521, USA (email: xinping.cui@ucr.edu). This work was partially supported by UC Riverside AES-CE RSAP A01869.

\newpage
\begin{center}
\textbf{Abstract}
\end{center}
The key of tip growth in eukaryotes is the polarized distribution on plasma membrane of a particle named ROP1. This distribution is the result of a positive feedback loop, whose mechanism can be described by a Differential Equation parametrized by two meaningful parameters $k_{pf}$ and $k_{nf}$. We introduce a mechanistic Integro-Differential Equation (IDE) derived from a spatiotemporal model of cell polarity and we show how this model can be fitted to real data i.e. ROP1 intensities measured on pollen tubes. At first, we provide an existence and uniqueness result for the solution of our IDE model under certain conditions. Interestingly, this analysis gives a tractable expression for the likelihood, and our approach can be seen as the estimation of a constrained nonlinear model. Moreover, we introduce a population variability by a constrained nonlinear mixed model. We then propose a constrained Least Squares method to fit the model for the single pollen tube case, and two methods, constrained Methods of Moments and constrained Restricted Maximum Likelihood (REML) to fit the model for the multiple pollen tubes case. The performances of all three methods are studied through simulations and are used on an in-house multiple pollen tubes dataset generated at UC Riverside.  

\vspace*{.3in}


\newpage

\section*{1. Introduction \label{section intro}}

Cell polarity is a fundamental feature of almost all cells. It is required for the differentiation of new cells, the formation of cell shapes, and cell migration, etc. Pollen tubes, which extend by an extreme form of polar growth (termed tip growth) to deliver sperms to the ovary for fertilization, are one of the fastest growing cells in plants and therefore represent an attractive model system to investigate polarized cell growth \citep{Yang1998,Hepler2001,LeeYang:2008,Yang:2008,QinYang:2011}.

When pollen grain is activated by a certain internal or external stimulus, the signaling molecule GTPase ROP1 in the cytosol will be activated and translocated onto the plasma membrane, forming an apical ROP1 cap.  Once maintained, the apical ROP1 cap will trigger exocytosis, leading to cell growth at the site of the apical cap. Figure 1 shows the three main stages of pollen tube tip growth: polarity establishment of the signaling molecule GTPase ROP1 (i.e, active ROP1s form a apical cap), exocytosis to increase cell membrane surface and deliver cell wall materials, and cell wall extension.  

\begin{figure}[ht]
\begin{center}
\includegraphics[trim = 0cm 6cm 0cm 6cm, clip, width=0.70\textwidth]{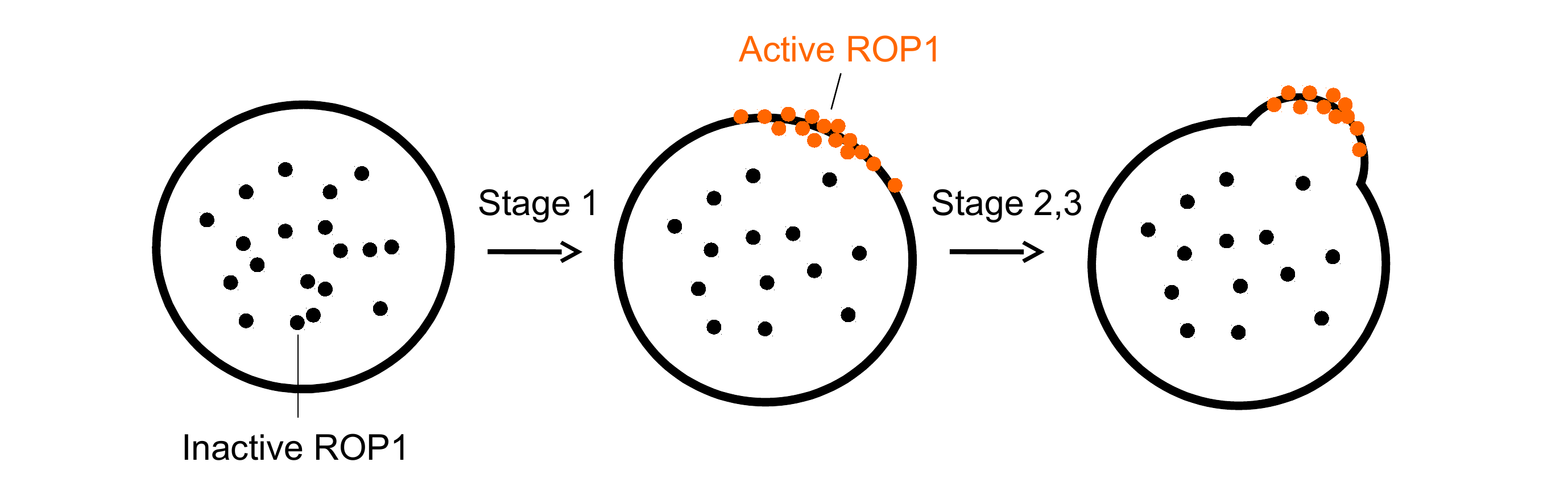}
\caption{Three main stages of tip growth of pollen tubes. Stage 1: ROP1 polarity establishment; Stage 2: Pectin Exocytosis; Stage 3: Cell wall extension. }
\end{center}
\end{figure}

Several mathematical models have been built to simulate pollen tube tip growth \citep{Dumais:2006,Kroeger2008,Campas2009,Lowery:2009,Fayant:2010}. These models focused on the cell wall mechanics and the cell wall mechanics-mediated shape formation of pollen tubes. However, it has been found that the generation of apical cap of active ROP1 at stage 1 plays a predominant role in determining polarity of the pollen tube \citep{Lin:1996,Li:1998}. Therefore, modeling the distribution of ROP1s on the membrane is the key to understand the tip growth of pollen tube. As a key regulator of the self-organizing pollen tube system, the activity and distribution of ROP1 are fine-tuned by both positive and negative feedback mechanisms \citep{Hwang:2010} as well as slow diffusion shown in Figure 2. \cite{Altschuler2008Spontaneous} proposed a linear differential equation model for the polarization of the GTPase Cdc42 in budding yeast but only considered positive feedback. On the other hand, for all the aforementioned models attention has been paid to predict or simulate the output using these models with given parameters. Less efforts have been devoted to the inverse problem, i.e., using the experimental data to estimate the parameters that characterize these models \citep{Ramsay2007,Wuhulin2008,Brunel2008,brunelhal00867370}.

In this paper, we propose an integro-ordinary differential equation (IDE) model to describe the three processes together (positive feedback, negative feedback and diffusion) that lead to ROP1 polarity formation at steady state. Our main interest lies in the inverse problem of estimating the parameters for the positive feedback and the negative feedback. However, two identifiability issues arise in the context of our model.  The first identifiability problem is whether the solution to the nonlinear IDE model exists and is unique. We will show that the IDE model is closely related to a semilinear elliptic equation, from which we establish the original theory on the existence and uniqueness of solutions to this type of IDE. The second identifiability problem is whether the observed data is enough to identify the parameters in the IDE model. By applying the identifiability analysis methods suggested by \cite{Miao:2011}, we can prove that the two parameters of interest are identifiable.

Solving the identifiability problems allows us to derive an admissible parameter space inside which the solution to the IDE model exists. The IDE model can then be re-parametrized as a mixed-effects differential equation model with linear constraints over the admissible parameter space. In statistical literature, there exist a number of papers for mixed-effects differential equation models. \cite{Li:2002} proposed to estimate time-varying parameters in the mixed-effects ordinary differential equations by maximizing the double penalized log likelihood. \cite{Putter:2002} proposed a hierarchical Bayesian approach for estimating population parameters in a system of mixed-effects nonlinear differential equations that have closed-form solutions. \cite{Guedj:2007} extended this system of mixed-effects nonlinear differential equations for which no closed form is available. They proposed to estimate both population and individual parameters in this extended model by a maximum likelihood approach using a Newton-like algorithm. \cite{Huang:2006a} and \cite{Huang:2006b} developed a hierarchical Bayesian approach to estimate both population and individual dynamic parameters in a set of mixed-effects nonlinear differential equations which have no closed-form solutions. \cite{Lu:2011} employed stochastic approximation EM approach for parameter estimation of mixed-effects ordinary differential equations. 

However, parameter estimation problems for mixed-effects differential equation models with linear inequality constraints have not been investigated. In this paper, we propose two algorithms based on modified REML and Method of Moments (MM) approaches \citep{1993}  to estimate the parameters with constraints in a mixed-effects differential equation model. The constrained estimators are shown to be consistent and the methodology we propose is quite general and can be applied to many mixed-effects ODE settings with little modification. 

The paper is organized as follows. In section 2, we introduce the IPDE and IDE model motivated by the GTPbase ROP1 polarization process. In the next section, we give sufficient and necessary conditions for existence and uniqueness of a positive solution to the IDE model, and we derive a tractable generic expression for solutions of the IDE. In section 4, we introduce the IDE based nonlinear statistical model with linear constraint for a single pollen tube.  We then extend the model for multiple pollen tubes and re-parametrize it as a nonlinear mixed model with linear constraints. The two estimators of Constrained Method of Moments (CMM) and Constrained REML(CREML) are proposed and the asymptotic properties of CMM are discussed. We examine the performance of the proposed estimation procedures through simulation studies in section 5 and real data analysis in section 6. We conclude the paper in section 7.

\section*{2. An Integral-Differential Equation Model of Cell Polarity}
To build the cell-signaling model of ROP1 polarity formation and maintenance, we assume that the redistribution of signaling molecules is determined by three fundamental transport mechanisms including (1) A positive feedback loop with rate $k_{pf}$ mediated by exocytosis and ROP1 activators such as RopGEFs \citep{Kost:1999,Li:1999,Berken:2005,Gu:2005,Lee:2008,McKenna:2009}; (2) A global negative regulation with rate $k_{nf}$ mediated by cytosolic ROP1 inhibitors such as RopGAPs \citep{Hwang:2008}; (3) Slow lateral diffusion of ROP1 protein on apical plasma membrane with rate $D$. 
These three processes are shown in Figure 2. The following semilinear Integro-Partial Differential Equation describes 
how these three processes together lead to ROP1 polarity formation:

\begin{figure}[ht]
\begin{center}
\includegraphics[trim = 0cm 3cm 6.5cm 6.5cm, clip, width=0.70\textwidth]{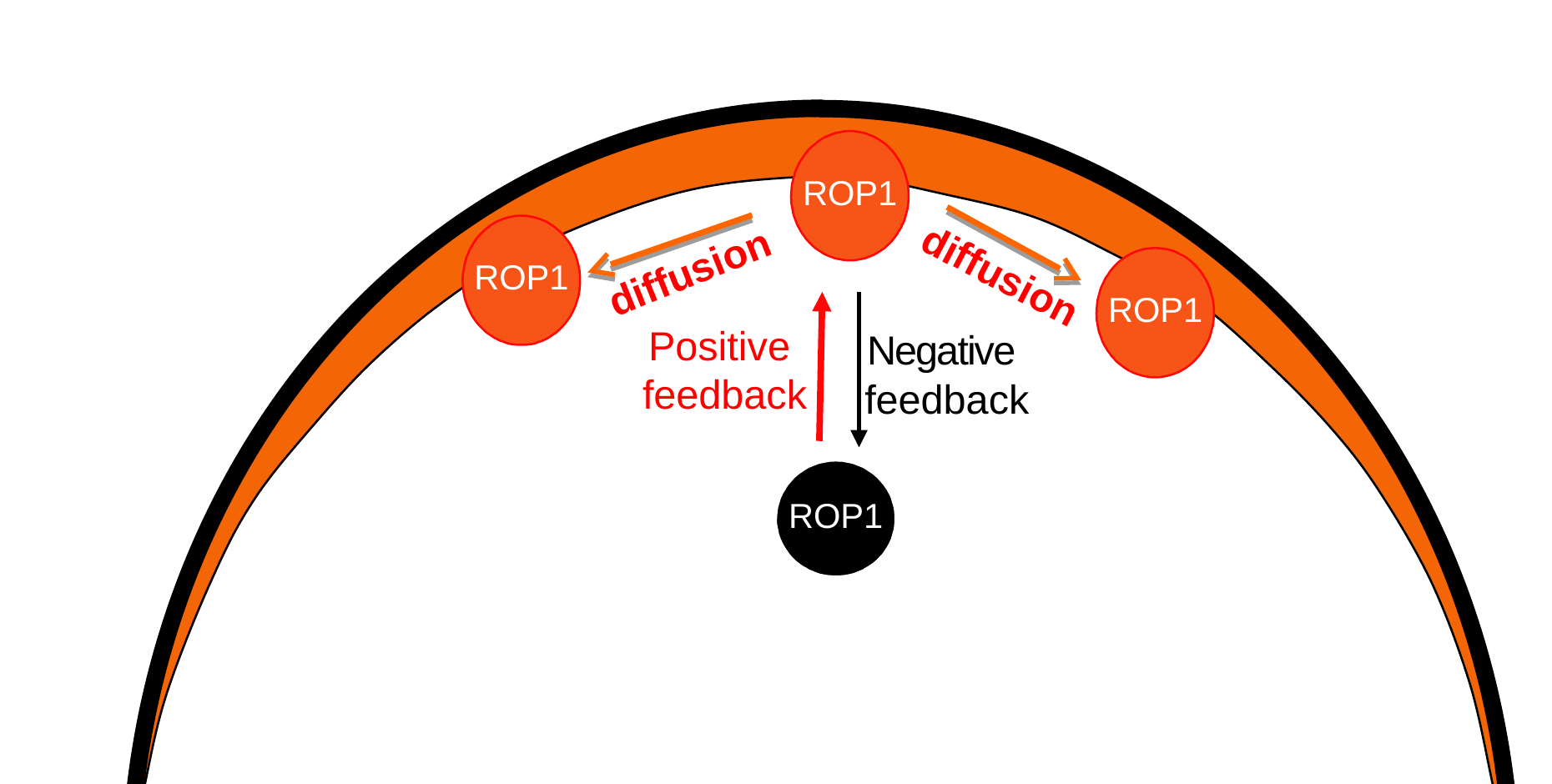}
\caption{ROP1 polarity formation is determined by positive feedback, negative feedback and lateral diffusion process.}
\end{center}
\end{figure}

\begin{equation}
\left\{
\begin{array}{l}
 \frac{\partial R(x,t)}{\partial t}= k_{pf}R(x,t)^{\alpha}(1-\frac{\int_{-L_0}^{L_0}{R(x,t) dx}}{R_{tot}})-k_{nf}R(x,t)+D\nabla^2 R(x,t) \\
 \text{where } \qquad \{x,t\}\in[-L_0,L_0]\times[0,\infty], \qquad R(-L_0,t)=R(L_0,t)=0
\end{array} \right. \label{eq:PDE}
\end{equation}                                   
$R(x,t)$ denotes the ROP1 intensity in position $x$ on the membrane at time $t$, which can be observed at a oblique plane of total length $2L_0$  passing through the cell center. $R_{tot}$ denotes the total free ROP1 in the cell. Throughout this paper,  $R_{tot}$, $D$, $\alpha>1$ and $L_0$ are assumed to be known constants. This model is similar as the PDE model described in \cite{Altschuler2008Spontaneous} except in their model spontaneous association was included, the fraction of all particles on the membrane is specified as $\frac{\int_{-L_0}^{L_0}{R(x,t) dx}}{R_{tot}}$ and $\alpha$ was assumed to be 1.

At equilibrium $t_0$, $\frac{\partial R(x,t)}{\partial t} |_{t=t_0}= 0$, $i.e.$, the ROP1 density won't change with time. From now on, we write $R(x,t_0)$ as $R(x)$. The IPDE model (\ref{eq:PDE}) then degenerates to the following IDE model  
\begin{equation}
\left\{
\begin{array}{ll}
-D \partial_{x}^{2}R=-k_{nf} R+k_{pf}R^{\alpha}(1-\frac{\int_{-L_0}^{L_0}{R dx}}{R_{tot}}) \\
\text{where } \qquad x\in[-L_0,L_0], \qquad R(-L_0)=R(L_0)=0
\end{array} \right. \label{eq:IDE}
\end{equation} 
Our interest lies in the estimation of the parameters for $k_{pf}$ and $k_{nf}$. However, two issues immediately arise, namely, the existence and uniqueness of the solution $R$ to the equation (2) and the identifiability of $k_{pf}$ and $k_{nf}$. Section 3 is devoted to address these two issues.

\section*{3. Identifiability Analysis}
In this section, we first show that the solution $R$ to the equation (2) exists and is unique. To the best of our knowledge, this is the first study on the existence and uniqueness of solution to integral differential equations when $\alpha \geq 1$. We then apply the identifiability analysis suggested by \cite{Miao:2011} and verified that parameters $k_{pf}$ and $k_{nf}$ are identifiable. 
\subsection*{3.1 Existence and Uniqueness of Solution $R$} 

\begin{lemma}
For all $c \in (0,\infty]$, there exists an unique positive solution $\sigma_0(x)$ to (\ref{eq:SemiEllipticEqu}) with Dirichlet conditions on $[-c,c]$.
\begin{equation}
  \left\{ \begin{array}{ll}
-\partial_{x}^{2}u=-u+u^{\alpha} & x\in[-c,c] \\
u(-c)=u(c)=0
\end{array}\right.\label{eq:SemiEllipticEqu}
\end{equation}
Moreover, $\sigma_0(x)$ is positive, even and increasing at $[-c,0]$ and decreasing at $[0,c]$.
\end{lemma}

\begin{lemma}
For all $\lambda^{'}>0$, there exists an unique positive solution $R_{\lambda'}(x)$ to \text{(\ref{eq:ParametrizedSemilinearElliptic})} with Dirichlet conditions on $\Omega=\left[-L_0,L_0\right]$, where $L_0 \in (0,\infty]$. 

\begin{equation}
\left\{ \begin{array}{ll}
-D\partial_{x}^{2}u=-k_{nf} u+\lambda' k_{pf} u^{\alpha}\\
\text{where } \qquad x\in[-L_0,L_0], \qquad u(-L_0)=u(L_0)=0
\end{array}\right.\label{eq:ParametrizedSemilinearElliptic}
\end{equation}  
Moreover, if $\sigma_{0}(x)$ is the unique positive solution to (\ref{eq:SemiEllipticEqu}) defined on $\Omega'=\left[-L_0\sqrt{\frac{k_{nf}}{D}},L_0\sqrt{\frac{k_{nf}}{D}}\right]$,
then 
\begin{equation}
R_{\lambda'}(x)=\left(\frac{k_{pf}}{k_{nf}}\lambda'\right)^{\frac{1}{1-\alpha}}\sigma_{0}(\sqrt{\frac{k_{nf}}{D}}x)\doteq R_{\lambda,\mu}(x)=\lambda \sigma_0(\mu x).\label{eq:ExpressionR_lambda}
\end{equation}
Where $\lambda=(\frac{k_{pf}}{k_{nf}} \lambda^{'})^{\frac{1}{1-\alpha}}$, and $\mu=\sqrt{\frac{k_{nf}}{D}}$.
\end{lemma}

The proofs of Lemma 1 and 2 are provided in the Appendix. It is easy to see that if there exists a unique positive solution $R_{\lambda'}(x)$ to equation (\ref{eq:ParametrizedSemilinearElliptic}) such that $\lambda^{'} =1-\frac{\int_x R_{\lambda'}(x)dx}{R_{tot}}$ , then $R_{\lambda'}(x)$ is also a solution to equation (\ref{eq:IDE}). In the following, we provide sufficient and necessary conditions that solutions to equation (\ref{eq:ParametrizedSemilinearElliptic}) are also solutions to equation (\ref{eq:IDE}).
 
\textbf{Sufficient Condition:} Let $\sigma_{0}$ be the positive solution to (\ref{eq:SemiEllipticEqu}) on $\Omega'=\left[-L_0\sqrt{\frac{k_{nf}}{D}},L_0\sqrt{\frac{k_{nf}}{D}}\right]$. Consider the family of function $R_{\lambda,\mu}(x)$ defined in (\ref{eq:ExpressionR_lambda}) and the discriminant function 
\begin{equation}
\Lambda(k_{nf},k_{pf},D,R_{tot},\sigma_{0})=\frac{k_{nf}}{k_{pf}}-\frac{1}{\alpha}\left(\frac{\alpha-1}{\alpha}\sqrt{\frac{k_{nf}}{D}}\frac{R_{tot}}{\left\Vert \sigma_{0}\right\Vert _{1}}\right)^{\alpha-1}\label{eq:ParameterCondition}
\end{equation}
If $\Lambda(k_{nf},k_{pf}, D,R_{tot},\sigma_{0})=0 (<0)$, then one (two) positive solution(s) to (\ref{eq:IDE}) can be found in the family of function $R_{\lambda,\mu}(x)$.

\textbf{Necessary Condition:} Any positive solution to (\ref{eq:IDE}) can be written in the form $R_{\lambda,\mu}(x)=\lambda\sigma_{0}(\mu x)$.

\begin{remark}
The proofs of sufficient condition and necessary condition are provided in the Appendix. As a result, the solution $R$ to (\ref{eq:IDE}) can be obtained as following when the values of $k_{nf}$ and $k_{pf}$ are given
\begin{enumerate}
\item Solve the semilinear elliptic equation (\ref{eq:SemiEllipticEqu}) on $\Omega'=\left[-L_0\sqrt{\frac{k_{nf}}{D}},L_0\sqrt{\frac{k_{nf}}{D}}\right]$.
\item Compute $\left\Vert \sigma_{0}\right\Vert _{1}$ and the discriminant function $\Lambda(k_{nf},k_{pf},D,R_{tot},\sigma_{0})$. 
\item If $\Lambda(k_{nf},k_{pf},D,R_{tot},\sigma_{0}) = 0$, find the positive roots $\lambda^*$ of $g(\lambda)$ (defined in the Appendix section A.3), and compute the solution $R_{\lambda^*,\mu}(x)$.
\item  If $\Lambda(k_{nf},k_{pf},D,R_{tot},\sigma_{0}) < 0$, find the positive roots $\lambda_1^*$ and $\lambda_2^*$ of $g(\lambda)$ (defined in the Appendix section A.3), and compute the solutions $R_{\lambda_1^*,\mu}(x)$ and $R_{\lambda_2^*,\mu}(x)$.
\end{enumerate}\label{Remark:1}
In practice, the solution closer to the experimental data should be chooen if there are two solutions.
\end{remark}

\begin{remark}
For $\lambda>0$, the solution $R_{\lambda,\mu}(x)$ to (\ref{eq:IDE}) is a positive and even function. Moreover, it increases at $\left[-L_0,0\right]$ and decreases at $\left[0,L_0\right]$, and the maximum $R_{\lambda,\mu}(0)=\max_{x \in \Omega}R_{\lambda,\mu}(x)>\lambda$. The proof is provided in the Appendix. \label{Remark:2} It will be shown later in section 7 that the ROP data reflects these qualitative properties.
\end{remark}

\begin{remark}
From sufficient condition, we see that the solution of (\ref{eq:IDE}) is determined by two parameters, $\mu$ and $\lambda$, which can be seen as a reparametrization of $k_{nf}$ and $k_{pf}$. Hence, (\ref{eq:IDE}) is not over parametrized by $k_{nf}$ and $k_{pf}$. \label{Remark:3}
\end{remark} 

\subsection*{3.2. Identifiability of $k_{nf}$ and $k_{pf}$} 

In this section, we prove that the parameters of interest $k_{nf}$ and $k_{pf}$ are globally identifiable and therefore are locally strongly identifiable. 

Let $R(x)$ denote the solution to (\ref{eq:IDE}).  In practice, $R(x)$ is a positive and non-constant function on interval $[-L_{0},L_{0}]\subset[-L,L]$. Suppose for parameters $(k_{nf}^{0},k_{pf}^{0})$ and $(k_{nf}^{1},k_{pf}^{1})$, $R(x;k_{nf}^{0},k_{pf}^{0})=R(x;k_{nf}^{1},k_{pf}^{1})$ on $[-L_{0},L_{0}]$, then we have
\begin{equation*}
k_{nf}^{0}R(x)-k_{pf}^{0}R^{\alpha}(x)\left(1-\frac{\int_{-L}^{L}R(x)dx}{R_{tot}}\right)=k_{nf}^{1}R(x)-k_{pf}^{1}R^{\alpha}(x)\left(1-\frac{\int_{-L}^{L}R(x)dx}{R_{tot}}\right)
\end{equation*}

For $R(x)>0$ on $[-L_{0},L_{0}]$
\begin{equation*}
k_{nf}^{0}-k_{nf}^{1}=(k_{pf}^{0}-k_{pf}^{1})R^{\alpha-1}(x)\left(1-\frac{\int_{-L}^{L}R(x)dx}{R_{tot}}\right)
\end{equation*}

If $k_{pf}^{0}-k_{pf}^{1}\ne{}0$, then
\begin{equation*}
R^{\alpha-1}(x)=(k_{nf}^{0}-k_{nf}^{1})/\left((k_{pf}^{0}-k_{pf}^{1})\left(1-\frac{\int_{-L}^{L}R(x)dx}{R_{tot}}\right)\right),
\end{equation*}
suggesting that $R(x)$ has to be constant on $[-L_{0},L_{0}]$ since $1-\frac{\int_{-L}^{L}R(x)dx}{R_{tot}}>0$. By contradiction, we can show that $k_{pf}^{0}-k_{pf}^{1}=0$ and $k_{nf}^{0}-k_{nf}^{1}=0$, i.e., $R(x;k_{nf}^{0},k_{pf}^{0})=R(x;k_{nf}^{1},k_{pf}^{1})$ on $[-L_{0},L_{0}]$ if and only if $k_{pf}^{0}=k_{pf}^{1}$ and $k_{nf}^{0}=k_{nf}^{1}$.

\section*{4. Parameter Estimation}
In this section, we first consider estimating $k_{nf}$ and $k_{pf}$ in a constrained nonlinear fixed effect model using a single pollen tube data. We then further extend to estimate $k_{nf}$ and $k_{pf}$ in a constrained nonlinear mixed effect model using multiple pollen tube data. 

\subsection*{4.1 Single pollen tube and constrained nonlinear fixed effect model}
Suppose for a single pollen tube, an observation of ROP1 intensity in a position $X_j$ ($X_j$ is randomly sampled from known distribution $F(x)$ such as an uniform distribution) on the membrane at equilibrium $t_0$ is denoted by
\begin{equation}
Y_j=R(X_j; k_{nf}, k_{pf})+\epsilon_j \qquad j=1,2,\cdots,n. \label{model:single1}
\end{equation}
where $R(X;\cdot)$ is the solution of (\ref{eq:IDE}) and $\epsilon_j$ are $iid$ from a certain distribution $f$ with mean 0 and variance $\sigma^2$. As shown in section 2, $R(X; \cdot)$ exists if and only if the discriminant function $\Lambda(\cdot)$ is non-positive. Therefore, the IDE based model (\ref{model:single1}) is subject to the constraint
\begin{equation*}
\left\{
\begin{array}{l}
\Lambda(k_{nf},k_{pf},D,R_{tot},\sigma_{0})=\frac{k_{nf}}{k_{pf}}-\frac{1}{\alpha}(\frac{\alpha-1}{\alpha}\sqrt{\frac{k_{nf}}{D}}\frac{R_{tot}}{\left\Vert \sigma_{0}\right\Vert _{1}})^{\alpha-1} \leq 0 \\
k_{nf} > 0, k_{pf} > 0
\end{array}
\right.
\end{equation*}

\begin{proposition}
The constrained nonlinear model (7) can be reparametrized into the following model (\ref{model:single2}) with $\mu$ and $\lambda$ subject to the constraint (\ref{Sec4Constraint})
\begin{equation}
Y_j=\lambda \sigma_0(\mu X_j)+\epsilon_j.\label{model:single2}
\end{equation}
\begin{equation}
\left \{
\begin{array}{l}
\Lambda^*(\mu,\lambda)=\mu R_{tot}-\lambda\left\Vert \sigma_{0}\right\Vert _{1} > 0 \\
\mu > 0, \lambda > 0
\end{array} \right. \label{Sec4Constraint}
\end{equation}
where $\mu=\sqrt{\frac{k_{nf}}{D}}$ and $\lambda$ is the root of $g(\lambda)$ (defined in the Appendix section A.3). The choice of $\lambda$ is discussed in Remark 1.
\end{proposition}
The proof of proposition 1 is provided in the Appendix. As a result, given the observations $\{y_j\}_{j=1}^{n}$ at positions $\{x_j\}_{j=1}^{n}$ from the biological experiment, we propose the following estimation method called Constrained Nonlinear Least Square (CNLS) method.
\begin{enumerate}
  \item Compute $\sigma_0(x)$ from DE (\ref{eq:SemiEllipticEqu})
  \item Estimate $\mu$ and $\lambda$ by minimizing (\ref{Sec4leastsquare}) under the constraint (\ref{Sec4Constraint})
	  \begin{equation}
      (\hat{\lambda},\hat{\mu})=\arg\min_{\lambda,\mu}\sum_{j=1}^{n}\left(y_{j}-\lambda\sigma_{0}(\mu x_{j})\right)^{2} \label{Sec4leastsquare}
    \end{equation}
  \item Convert $\hat{\mu}$ and $\hat{\lambda}$ to $\hat{k}_{nf}$ and $\hat{k}_{pf}$ by $\hat{k}_{nf}=D\hat{\mu}^2$ and $\hat{k}_{pf}=\frac{D\hat{\mu}^2}{\hat{\lambda}^{\alpha-1}-\frac{\hat{\lambda}^{\alpha}\left\Vert\sigma_{0}\right\Vert _{1}}{\hat{\mu}R_{tot}}}$  
    \item Estimate $\sigma^2$ by $\hat{\sigma}^2=\frac{1}{n}\sum_{j=1}^{n}\left(y_{j}-\hat{\lambda}\sigma_{0}(\hat{\mu} x_{j})\right)^{2}$
\end{enumerate}

In the first step of CNLS, the solution of $\sigma_0$ involves a boundary value problem in an ordinary differential equation, which can be solved by many methods including shooting method \citep{Soetaert,Soetaert:Petzoldt:Setzer:2010:JSSOBK:v33i09}, mono-implicit Runge-Kutta (MIRK) method \citep{Cash2005362} and collocation method \citep{Bader:1987:NBI:36006.36008} in R package ``bvpSolve''. The optimization in the second step is subject to one linear constraint and two box constraints. When there is no constraint, the optimization can be tackled by many gradient based methods which require the objective function to be differentiable, such as the Newton method, the BFGS method, the Gauss-Newton method, etc. On the other hand, the simplex method \citep{citeulike:3009487} that directly searches the optimum allows the objective function to be not differentiable. To apply the simplex method, we first incorporate the constraints into the objective function by defining
\begin{equation*}
f(\mu,\lambda)= \left \{
\begin{array}{cl}
 \sum_{j=1}^{n}\left(y_{j}-\lambda\sigma_{0}(\mu x_{j})\right)^{2} & \text{if  } \Lambda^*(\mu,\lambda) > 0 \text{  and  } \mu>0 \text{  and  } \lambda>0\\
  +\infty &  o.w.
\end{array} \right.
\end{equation*} 
The same idea was used by \cite{citeulike:3009487}.

For the general Nonlinear Least Square (NLS) estimator, the asymptotic properties have been established by \cite{citeulike:12197486}. For the general Constrained NLS (CNLS) estimator, the asymptotic properties have been established by \cite{citeulike:1392995}. Below we present the asymptotic properties of the CNLS estimator proposed in this paper. The proof is provided in the Appendix.

\begin{proposition}
Let $\boldsymbol{\theta}=(\mu,\lambda)^T$ be the parameter vector, $\boldsymbol{\theta}_0$ be the true value of $\boldsymbol{\theta}$, and $\hat{\boldsymbol{\theta}}_n$ be the CNLS estimator with $n$ sample points. Let $R(X;\boldsymbol{\theta})=\lambda \sigma_0(\mu X)$, then 
\begin{equation*}
\sqrt{n}(\hat{\boldsymbol{\theta}}_n-\boldsymbol{\theta}_0)\stackrel{d}{\rightarrow} N(0,\sigma^2 K^{-1}) 
\end{equation*} 
where $K=E_{X}[\nabla_{\boldsymbol{\theta}}R(X;\boldsymbol{\theta}_0)\nabla_{\boldsymbol{\theta}}R(X;\boldsymbol{\theta}_0)^T]$, and $\nabla_{\boldsymbol{\theta}}R(X;\boldsymbol{\theta}_0)$ is the gradient vector of $R(X;\boldsymbol{\theta})$ with respect to $\boldsymbol{\theta}$ at $\boldsymbol{\theta}=\boldsymbol{\theta}_0$. 
\end{proposition}

\begin{proposition}
Let the estimate of $\sigma^2$ be $\hat{\sigma}_{n}^2=\frac{1}{n}\sum_{j=1}^{n}\left(y_{j}-\hat{\lambda}\sigma_{0}(\hat{\mu} x_{j})\right)^{2}$. Then
\begin{equation*}
\hat{\sigma}_{n}^2\stackrel{p}\rightarrow \sigma^2
\end{equation*}
Then by \textit{Slutsky's} Theorem
 \begin{equation*}
   \frac{\sqrt{n}(\hat{\boldsymbol{\theta}}_n-\boldsymbol{\theta}_0)}{\hat{\sigma_n}} \stackrel{d}{\rightarrow}  N(0,K^{-1})
\end{equation*}
\end{proposition}

\begin{corollary}
Denote $\boldsymbol{\phi}=(k_{nf},k_{pf})^T$. Let $\boldsymbol{\phi}_0$ and $\hat{\boldsymbol{\phi}}_n$ be the true value and estimator of $\boldsymbol{\phi}$ respectively, where $\hat{k}_{nf}=D\hat{\mu}^2$ and $\hat{k}_{pf}=\dfrac{D\hat{\mu}^2}{\hat{\lambda}^{\alpha-1}-\dfrac{\hat{\lambda}^\alpha\parallel\sigma_0\parallel_1}{\hat{\mu}R_{tot}}}$. By the delta-method,   

 \begin{equation*}
   \frac{\sqrt{n}(\hat{\boldsymbol{\phi}}_n-\boldsymbol{\phi}_0)}{\hat{\sigma_n}} \stackrel{d}{\rightarrow}  N(0,A^TK^{-1}A) 
 \end{equation*}
where
 \begin{equation*}
A=\left( \begin{array}{cc}
          \frac{\partial k_{nf}}{\partial \mu} &  \frac{\partial k_{pf}}{\partial \mu}\\
          \frac{\partial k_{nf}}{\partial \lambda} &  \frac{\partial k_{pf}}{\partial \lambda}
          \end{array}
   \right)=\left( \begin{array}{cc}
          2D\mu &  \frac{2D\mu^3-3D\mu^2\lambda\frac{\left\Vert \sigma_{0}\right\Vert _{1}}{R_{tot}}}{\lambda^{\alpha-1}(\mu-\lambda\frac{\left\Vert \sigma_{0}\right\Vert _{1}}{R_{tot}})^2} \\
          0 & \frac{-D\mu^4(\alpha-1)+D\mu^3\alpha\lambda\frac{\left\Vert \sigma_{0}\right\Vert _{1}}{R_{tot}}}{\lambda^{\alpha}(\mu-\lambda\frac{\left\Vert \sigma_{0}\right\Vert _{1}}{R_{tot}})^2}
          \end{array}
   \right)
 \end{equation*} 
\end{corollary}

\subsection*{4.2 Multiple pollen tubes and constrained nonlinear mixed effect model}

In this section, we consider multiple pollen tubes and extend model \ref{model:single1} and \ref{model:single2} as follows: 
\begin{eqnarray}
 Y_{ij}&=&R_i(X_{ij};\lambda_i,\mu_i)+\epsilon_{ij} \\
       &=&\lambda_i\sigma_0(\mu_i X_{ij})+\epsilon_{ij}, \qquad i=1,2,\cdots, m; j=1,2,\cdots,n_i,
\end{eqnarray}
where $Y_{ij}$ denotes the ROP1 intensity observed for pollen tube $i$ at position $X_{ij}$ on the membrane at static time and $\epsilon_{ij}$ is $i.i.d$ with distribution $N(0,\sigma^2)$. We further assume that

\begin{equation} \label{eq:21}
\begin{cases}
(\mu_i, \lambda_i)^T=(\mu, \lambda)^T+\Phi_i\\
\lambda_i>0, \quad \mu_i>0
\end{cases}
\end{equation}
where $\Phi_i$ is $i.i.d$ with distribution $MVN(0,\Sigma)$. As a result, this is a nonlinear mixed model (NMM) where $\sigma^2$ measures within pollen tube variation and $\Sigma$ measures between pollen tube variation. As discussed in Section 4.1, each pair of $\lambda_i>0, \mu_i$ are subject to three constraints. If no constraint exists, all the parameters can be estimated by several existing methods such as \cite{Wangyuedong} and \cite{Wolfinger}.

\paragraph{}
Denote $\boldsymbol{\theta}_i=(\mu_i,\lambda_i)^T$ and $\boldsymbol{\theta}=(\mu,\lambda)^T$, and the experimental data to be $\{y_{ij}\}$ and $\{x_{ij}\}$ with $i=1,\cdots,m$ and $j=1,\cdots,n_i$. We first extend the CNLS procedure and propose a new procedure called Constrained Method of Moment (CMM) as follows:
\begin{enumerate}
  \item Compute $\sigma_0(x)$ from equation (\ref{eq:SemiEllipticEqu})
	\item For each pollen tube $i$, estimate $\boldsymbol{\theta}_i$ by minimizing least squares
        \begin{equation*}
          \hat{\boldsymbol{\theta}}_i=\arg\min_{\boldsymbol{\theta}_i}\sum_{j=1}^{n_i}\left(y_{ij}-\lambda_i\sigma_{0}(\mu_i x_{ij})\right)^{2}
        \end{equation*}
       under the constraints $\Lambda^*(\boldsymbol{\theta}_i) > 0$ and $\boldsymbol{\theta}_i>0$
	\item Estimate $\boldsymbol{\theta}$ by $\hat{\boldsymbol{\theta}}=\frac{\sum_{i=1}^m \hat{\boldsymbol{\theta}}_i}{m}$
	\item Estimate $\sigma^2$ by $\hat{\sigma}^2=\frac{\sum_{i=1}^{m}\sum_{j=1}^{n_i}\left(y_{ij}-\hat{\lambda}_i\sigma_{0}(\hat{\mu}_i x_{ij})\right)^{2}}{\sum_{i=1}^m (n_i-2)}$
	\item Estimate $\Sigma$ by $\hat{\Sigma}=\sum_{i=1}^m\frac{(\hat{\boldsymbol{\theta}}_i-\hat{\boldsymbol{\theta}})( \hat{\boldsymbol{\theta}}_i-\hat{\boldsymbol{\theta}})^T}{m-1}-\hat{\sigma}^2\sum_{i=1}^m \frac{T_i^{-1}}{m}$, where $T_i=\left[ \frac{\partial \boldsymbol{R}_i}{\partial \boldsymbol{\theta}_i^T}\right]^T\left[ \frac{\partial \boldsymbol{R}_i}{\partial \boldsymbol{\theta}_i^T} \right] \bigg|_{\boldsymbol{\theta}_i=\hat{\boldsymbol{\theta}}_i}$ and $\boldsymbol{R}_i=(R(x_{i1};\boldsymbol{\theta}_i), R(x_{i2};\boldsymbol{\theta}_i),\cdots,R(x_{in_i};\boldsymbol{\theta}_i) )^T$
	\item Modify the estimator of $\Sigma$ by 
	    \begin{equation*}
         \Sigma^* = \left\{
         \begin{array}{ll}
           \hat{\Sigma} & \text{if  }  \hat{\Sigma} \text{  is positive definite}\\
           \hat{\Sigma}_+ & \text{if  }  \hat{\Sigma} \text{  is not positive definite}
         \end{array} \right.
      \end{equation*}
where $\hat{\Sigma}_+ =Q\Psi_+ Q'$, in which $\Psi_+$ is a diagonal matrix whose diagonal elements $\Psi_{ii}= max(\psi_i, 0)$ where $\psi_i$ is the eigenvalue of $\Sigma$, and Q is a $2\times 2$ matrix whose $i^{th}$ columns is the eigenvector $q_i$ associated with $\psi_i$.     
	\item Convert $\hat{\boldsymbol{\theta}}$ to $\hat{k}_{nf}$ and $\hat{k}_{pf}$
\end{enumerate}
This procedure is motivated by the Method of Moments (MM) proposed by \cite{1993}. Our contribution is to extend it to constrained case by adding a constraint in the second step. Below we establish the asymptotic properties of the CMM estimators $\hat{\theta}$, $\Sigma_\Phi$. The proof is provided in the Appendix.

\begin{proposition}
Assume that 
\begin{enumerate}
  \item the sample size from each pollen tubes are equal, i.e., $n_1=n_2=\cdots=n_m=n$
	\item both $n$ and $m$ tend to $+\infty$
\end{enumerate}
Then, we have the following large sample properties for $\hat{\boldsymbol{\theta}}$ 
\begin{enumerate}
	\item $\hat{\boldsymbol{\theta}} \stackrel{p}{\rightarrow} \boldsymbol{\theta}$
	\item $\sqrt{m}\tilde{\Sigma}^{-\frac{1}{2}}(\hat{\boldsymbol{\theta}}- \boldsymbol{\theta}) \stackrel{d}{\rightarrow} \boldsymbol{Z}$, where $\boldsymbol{Z} \sim N(0,I_2)$, and $\tilde{\Sigma}=\Sigma + \sigma^2 E_{\boldsymbol{\theta}}[(nK_i)^{-1}]$ \\
          with $K_i=E_{X}[\nabla_{\boldsymbol{\theta}_i}R(X;\boldsymbol{\theta}_i)\nabla_{\boldsymbol{\theta}_i}R(X;\boldsymbol{\theta}_i)^T]$.
\end{enumerate}
Moreover, if $\hat{\sigma}^2$ is a consistent estimator of $\sigma^2$, then  
\begin{enumerate}
	\item $\hat{\Sigma} \stackrel{p}{\rightarrow} \Sigma$
	\item $\sqrt{m}\hat{\tilde{\Sigma}}^{-\frac{1}{2}}(\hat{\boldsymbol{\theta}}- \boldsymbol{\theta}) \stackrel{d}{\rightarrow} \boldsymbol{Z}$, where $\hat{\tilde{\Sigma}}=\hat{\Sigma} + \hat{\sigma}^2 E_{\boldsymbol{\theta}}[(nK_i)^{-1}]$.
\end{enumerate}
\end{proposition}

\begin{corollary}
Denote $\boldsymbol{\phi}=(k_{nf},k_{pf})^T$. Let $\boldsymbol{\phi}_0$ and $\boldsymbol{\hat{\phi}}$ be the true value and estimator of $\boldsymbol{\phi}$ respectively. By the delta-method,
 \begin{equation*}
   \sqrt{m}(A^T\hat{\tilde{\Sigma}}A)^{-\frac{1}{2}}(\hat{\boldsymbol{\phi}}-\boldsymbol{\phi}_0) \stackrel{d}{\rightarrow}  \boldsymbol{Z} 
 \end{equation*}
where $A$ is given previously. 
\end{corollary}

\paragraph{}

Note that the CMM requires the same sample size among subjects, which is usually violated in real data. If $\epsilon_{ij}$ are $iid$ normal, we can convert the nonlinear mixed model to a linear mixed model by Taylor approximation, and thereafter propose an alternative procedure called Constrained Restricted Maximum Likelihood method (CREML) as follows: 

\begin{enumerate}
  \item Given current Best Linear Unbiased Predictors (BLUP) $(\hat{\mu}_i^{(t)},\hat{\lambda}_i^{(t)})$ for $(\mu_i,\lambda_i)$, use Taylor expansion to express $R_i(\mu_i,\lambda_i;x)$ as
	     \begin{equation*}
	       R_i(\mu_i,\lambda_i;x) \approx R_i(\hat{\mu}_i^{(t)},\hat{\lambda}_i^{(t)};x)+\frac{\partial R_i}{\partial \mu_i}|_{\mu_i=\hat{\mu}_i^{(t)}}(\mu_i-\hat{\mu}_i^{(t)})+\frac{\partial R_i}{\partial \lambda_i}|_{\lambda_i=\hat{\lambda}_i^{(t)}}(\lambda_i-\hat{\lambda}_i^{(t)})
	     \end{equation*} 
	     As a result, the original expression of data $y_{ij}=R_i(\mu_i,\lambda_i; x_{ij})+\epsilon_{ij}$ can be re-written as
\begin{equation}
\begin{split}
	       y^*_{ij}&=\frac{\partial R_i}{\partial \mu_i}|_{\mu_i=\hat{\mu}_i^{(t)}}\mu_i+\frac{\partial  R_i}{\partial \lambda_i}|_{\lambda_i=\hat{\lambda}_i^{(t)}}\lambda_i+\epsilon_{ij}\\ \label{eq:LMM}  
          &= \left(\begin{array}{c}
\dfrac{\partial R_i}{\partial \mu_i}|_{\mu_i=\hat{\mu}_i^{(t)}} \\ \dfrac{\partial R_i}{\partial \lambda_i}|_{\lambda=\hat{\lambda}_i^{(t)}}
\end{array}\right)^T\left(\begin{array}{c} \mu \\ \lambda
\end{array}\right)+\left(\begin{array}{c}
\dfrac{\partial R_i}{\partial \mu_i}|_{\mu_i=\hat{\mu}_i^{(t)}} \\ \dfrac{\partial R_i}{\partial \lambda_i}|_{\lambda=\hat{\lambda}_i^{(t)}}
\end{array}\right)^T\Phi_i+\epsilon_{ij}
\end{split}
\end{equation}
where, $y^*_{ij}=y_{ij}-R_i(\hat{\mu}_i^{(t)},\hat{\lambda}_i^{(t)};x_{ij})+\frac{\partial R_i}{\partial \mu_i}|_{\mu_i=\hat{\mu}_i^{(t)}}\hat{\mu}_i^{(t)}+\frac{\partial R_i}{\partial \lambda_i}|_{\lambda_i=\hat{\lambda}_i^{(t)}}\hat{\lambda}_i^{(t)}$, $\Phi_i \overset{iid}\sim MVN(0,\Sigma)$,  $\epsilon_{ij} \overset{iid}\sim N(0,\sigma^2)$, $\Lambda^*(\mu, \lambda)=\mu R_{tot}-\lambda\parallel\sigma_0\parallel_1>0$, $\lambda>0$, $\mu>0$.  And our original model becomes a Constrained Linear Mixed Effect Model (CLMM).
	\item Fit CLMM (\ref{eq:LMM}) under the constraints of $\Lambda^*(\mu,\lambda) > 0$, $\mu>0$ and $\lambda>0$. Such constraints at the population level can be easily embraced by REML.
	\item Update $(\hat{\mu}_i^{(t)},\hat{\lambda}_i^{(t)})$ by the Best Linear Unbiased Predictors (BLUP) based on the Best Linear Unbiased Estimates (BLUE) $(\hat{\mu},\hat{\lambda},\hat{\Sigma},\hat{\sigma}^2)$ of the CLMM (\ref{eq:LMM}) from step 2.
	\item Iterate the above three steps until convergence.
\end{enumerate}
This procedure is motivated by the iterative procedure of \cite{1990}. Our contribution is to extend it to constrained case by adding a constraint on Step 2 and to use a simple way to update $(\hat{\mu}_i^{(t)},\hat{\lambda}_i^{(t)})$ in the iteration process.

\paragraph{}
The convergence behavior of the CREML procedure depends on the starting value. A good choice of starting value could be the estimates of the CMM procedure. If no constraints exist, the CLMM model in step 2 can be fitted by many existing approaches such as MLE, REML and EM algorithm. In this paper, we consider REML and extend it to fit the model with constraints. Note that the likelihood in the first step of REML only involves the variance component parameters $\Sigma$ and $\sigma^2$, therefore their estimates won't be affected by the constraints. On the other hand, the likelihood in the second step of REML involves the population parameters $\mu$ and $\lambda$. So their estimates should be obtained by maximizing the reduced likelihood under the constraints. And this constrained maximization problem was discussed in the previous section of single pollen tube case.

\paragraph{}
Note that the CMM procedure controls the constraints at the individual level whereas the CREML procedure controls them at the population level. 
Since constraints satisfied at the individual level will be automatically satisfied at the population level, the former is more strict than the latter. In many cases of real world application especially when the number of pollen tubes, $m$ is large, constraints at the population level is more desirable.

\section*{5. Simulation Studies}

In this section, simulation studies were conducted for the cases of single pollen tube and multiple pollen tubes respectively. All the estimation procedures were implemented in R. From the proof of Remark \ref{Remark:2}, we know $\sigma_0(x)$ is a positive and even function that achieves its maximum at 0. Further, we know $\sigma_0(x)$ is close to $\frac{1}{2}$ when $|x|=5$ and is close to 0 when $|x|\geq15$. Therefore, when $\mu=1$, $R(x)=\lambda\sigma_0(\mu x)$ is close to 0 when $|x|\geq15$. Therefore, in the simulation the data of $R(x)$ for $\mu=1$ were generated from $|x|<15$. The values of $\alpha$, $D$ and $R_{tot}$ used in the simulations were set to be $1.2$, $0.1$ and $797$ respectively, which were obtained empirically from real data.

\subsection*{5.1 Single pollen tube}

To evaluate the performance of the CNLS procedure, we simulated data based on Remark 1 using the true values $k_{nf}=0.1$, $k_{pf}=0.1125$.Therefore, $\mu=1$ and $\lambda=34.1883$. Since the range of $R(x)$ is $[0,55.06]$, we set the true value of measurement error $\sigma$ to be 4, 8, 16. For different $\sigma$, we generated 10000 data sets of size $n=301$, i.e., $x$ were picked along $[-15,15]$ with step size 0.1. CNLS based estimates of the parameters were obtained for each of the 10000 data sets, based on which the relative bias, standard deviation were computed as shown in Table 1. From Table 1, we could see the CNLS procedure works quite well and is quite robust against noise when the size of data is fairly large. We also followed Proposition 3 to compute asymptotical variances and construct the coverage probability as shown in Table 1. $K=E_{X}[\nabla_{\boldsymbol{\theta}}R(X;\boldsymbol{\theta}_0)\nabla_{\boldsymbol{\theta}}R(X;\boldsymbol{\theta}_0)^T]$ in Proposition 3 can not be computed analytically. However, when $n \geq 300$, it can be well approximated by its sample mean $\frac{1}{n}\sum_{i=1}^n\nabla_{\boldsymbol{\theta}}R(x_i;\boldsymbol{\theta}_0)\nabla_{\boldsymbol{\theta}}R(x_i;\boldsymbol{\theta}_0)^T$ according to our simulation. From Table 1, we could see that the asymptotical variances computed based on Proposition 3 are close to that computed based on simulation, and the observed coverage appears to be approximately equivalent to the nominal confidence level.
 
\begin{table}[htb]
\begin{center}
\begin{tabular}{ccccccc}
\hline
\multicolumn{2}{c}{} & $\hat{k}_{nf}$ & $\hat{k}_{pf}$ & $\hat{\mu}$ & $\hat{\lambda}$ & $\hat{\sigma}_\epsilon$ \\
\hline
\multirow{3}{*}{$\sigma=4$} & Bias & $-3.6*10^{-5}$ & $4.4*10^{-5}$ & $-2.8*10^{-4}$ & $1.0*10^{-2}$ & $-1.2*10^{-2}$\\ 
 & $sd$ & 0.0028 & 0.0022 & 0.0138 & 0.3963 & 0.1616 \\
 & $sd*$ & 0.0028 & 0.0023 & 0.0140 & 0.4071 & \\
& conv. prob. & 0.945 & 0.956 & 0.946 & 0.953 & \\
 \hline
 \multirow{3}{*}{$\sigma=8$} & Bias & 0.0001 & 0.0002 & 0.0002 & 0.0294 & -0.0479 \\
  & $sd$ & 0.0055 & 0.0046 & 0.0277 & 0.8151 & 0.3274 \\
  & $sd^*$ & 0.0056 & 0.0046 & 0.0279 & 0.8141 & \\
  & conv. prob. & 0.95 & 0.947 & 0.949 & 0.944 & \\
  \hline
  \multirow{3}{*}{$\sigma=16$} & Bias & 0.0004 & 0.0009 & 0.0008 & 0.1087 & -0.0417 \\
   & $sd$ & 0.0106 & 0.0087 & 0.0527 & 1.5477 & 0.6689 \\
   & $sd^*$ & 0.0111 & 0.0092 & 0.0559 & 1.6283 & \\
    & conv. prob. & 0.961 & 0.951 & 0.958 & 0.954 & \\
   \hline
\end{tabular}
\caption{CNLS estimators. sd: estimated standard deviation; $sd^*$: theoretical standard deviation based on proposition 3.}
\label{table:1}
\end{center}
\end{table}

\subsection*{5.2 Multiple pollen tubes}

To evaluate and compare the performance of the CMM and CREML procedures, we generated data for each $m=10$ pollen tubes based on Remark \ref{Remark:1} and associated $(\mu_i, \lambda_i)$ simulated from $MVN((\mu,\lambda)^T,\Sigma)$. The true values of parameters used for the simulation were $k_{pf}=0.1, k_{nf}=0.1125, \mu=1, \lambda=34.1883, \sigma=4$ and $\Sigma$ is a diagonal matrix with 
$\Sigma_{11}=0.04$ and $\Sigma_{22}=0.36$. We considered two cases. In case 1, $x=(-5,-1,-0.2,0.2,1,5)$ and $n=6$. In case 2, x is uniformly sampled from -5 to 5 with step size 0.2 and $n=51$. Each simulation was done 1000 times. The relative bias, standard deviation and coverage probability for CMM and CREML procedures are shown in Table \ref{table:3} and Table \ref{table:4}.

\begin{table}[htb]
\centering
\begin{tabular}{ccccccc}
\hline
\multicolumn{3}{c}{} & $\hat{k}_{nf}$ & $\hat{k}_{pf}$ & $\hat{\mu}$ & $\hat{\lambda}$ \\
\hline
\multirow{6}{*}{Case 1} & \multirow{5}{*}{CMM} & Bias & 0.0037 & 0.0021 & 0.0157 & 0.0135\\ 
& & sd & 0.0147 & 0.0068 & 0.0709 & 0.4871\\
& & $sd^*$ & 0.0139 & 0.0060 & 0.0695 & 0.4569\\ 
& & conv. prob. & 0.932 & 0.923 & 0.940 & 0.942 \\
\cline{2-7}
 & \multirow{2}{*}{CREML} & Bias & -0.0004 & 0.0002 & -0.0039 & -0.0204 \\
 & & sd & 0.0123 & 0.0053 & 0.0614 & 0.4573\\
  \hline
\multirow{6}{*}{Case 2} & \multirow{5}{*}{CMM} & Bias & 0.0016 & 0.0011 & 0.0061 & 0.0210\\ 
& & sd & 0.0123 & 0.0052 & 0.0607 & 0.2759\\
& & $sd^*$ & 0.0128 & 0.0053 & 0.0639 & 0.2728\\ 
& & conv. prob. & 0.961 & 0.949 & 0.960 & 0.945 \\
\cline{2-7}
 & \multirow{2}{*}{CREML} & Bias & 0.0015 & 0.0011 & 0.0059 & 0.0181 \\
 & & sd & 0.0122 & 0.0051 & 0.0606 & 0.2737\\
  \hline
\end{tabular}
\caption{Parameter estimation by CMM and CREML}
\label{table:2}
\end{table}

\begin{table}[htb]
\centering
\begin{tabular}{cccccc}
\hline
\multicolumn{3}{c}{} & $\hat{\sigma}$ & $\hat{\Sigma}_{11}$ & $\hat{\Sigma}_{22}$ \\
\hline
\multirow{4}{*}{Case 1} & \multirow{2}{*}{CMM} & Bias & 0.0155 & 0.0103 & 0.4409 \\ 
& & sd &  0.6050 & 0.0468 & 1.9676 \\
\cline{2-6}
 & \multirow{2}{*}{CREML} & Bias & -0.0741 & -0.0100 & 0.0974\\
 & & sd & 0.4147 & 0.0164 & 0.6516\\
  \hline
\multirow{4}{*}{Case 2} & \multirow{2}{*}{CMM} & Bias & -0.0021 & -0.0012 & 0.0227 \\ 
& & sd &  0.1312 & 0.0183 & 0.3383 \\
\cline{2-6}
 & \multirow{2}{*}{CREML} & Bias & -0.0039 & -0.0067 & -0.0513\\
 & & sd & 0.1305 & 0.0156 & 0.2942\\
  \hline
\end{tabular}
\caption{Variance components estimation by CMM and CREML }
\label{table:3}
\end{table}

From Table \ref{table:2} and Table \ref{table:3}, we can see that when $n$ is large, CMM procedure and the CREML procedure perform equally well. When $n$ is small, however, the CREML procedure performs better than the CMM procedure. Similar results were also observed by Munther Al-Zaid \cite{BIMJ:BIMJ881}. 

\section*{6. Pollen tube data analysis}

ROP1 intensities from 12 pollen tubes of Arabidopsis were collected at positions (-10$\mu m$, 10 $\mu m$) with step size 0.1205 $\mu m$.Therefore, $m=12$ and $n=173$. The ROP1 intensities in different pollen tubes are believed to have identical distributions. Therefore, quantile normalization was applied to normalize raw data and possible outliers were removed. Notice that the data of $R(x)$ is not $0$ even the images show no ROP intensity at $x$. Therefore, we pool the data sets together and fit the pooled data nonparametrically to obtain $\hat{R}(x)$, and set the background noise to be the smallest value of $\hat{R}(x)$. Then, subtract the background noise from $\hat{R}(x)$ and all the data points. We then standardize $\hat{R}(x)$ and all the data points to $\hat{R}(x)$ with range from 0 to 1 in order to get rid of the unit effects. The values for $D$, $R_{tot}$ and $\alpha$ used in the study were $0.2$, $30$ and $1.2$, respectively. 

We first performed CNLS procedure to the pooled normalized data sets and the individual data sets. The estimates of $k_nf$ and $k_pf$ for individual tubes are presented in table \ref{table:4} and for pooled data are 0.1930 and 0.2979, respectively. As we can see, the estimates from each individual tube are close to each other as well as to those obtained from pooled data. This is due to the fact that the sample size within each pollen tube is sufficiently large. Moreover, this indicates that the variation between pollen tubes is not too large. In addition, we performed CMM procedure and CREML procedure to the normalized data sets and the results are shown in table \ref{table:5}.

\begin{table}[htb]
\centering
\begin{tabular}{ccccccccc}
\hline
 & $\hat{k}_{nf}$ & $\hat{k}_{pf}$ & & $\hat{k}_{nf}$ & $\hat{k}_{pf}$ & & $\hat{k}_{nf}$ & $\hat{k}_{pf}$ \\
 \hline
 Tube 1 & 0.1866 & 0.2925 & Tube 2 & 0.2278 & 0.3337 & Tube 4 & 0.1814 & 0.2854 \\
 Tube 5 & 0.2205 & 0.3265 & Tube 6 & 0.1788 & 0.2748 & Tube 7 & 0.1892 & 0.2925 \\
 Tube 8 & 0.1917 & 0.2977 & Tube 9 & 0.2121 & 0.3188 & Tube 10 & 0.1976 & 0.3053 \\
 Tube 11 & 0.1939 & 0.3011 & Tube 14 & 0.1694 & 0.2766 & Tube 15 & 0.1809 & 0.2810 \\
 \hline
\end{tabular}
\caption{Results of CNLS procedure for individual tube}
\label{table:4}
\end{table}

\begin{table}[htb]
\centering
\begin{tabular}{ccccccccc}
\hline
 & $\hat{k}_{nf}$ & $\hat{k}_{pf}$ & $\hat{\mu}$ & $\hat{\lambda}$ & $\hat{\sigma}_\epsilon$ & $\hat{\sigma}_\mu$ & $\hat{\sigma}_\lambda$ & $\hat{\rho}_{\mu,\lambda}$ \\ 
 \hline
 CMM & 0.1942 & 0.2987 & 0.9708 & 0.6477 & 0.2064 & 0.0789 & 0.0393 & 0.737 \\
 CREML & 0.1862 & 0.2873 & 0.9648 & 0.6487 & 0.2267 & 0.0709 & 0.0258 & 0.838\\
 \hline
\end{tabular}
\caption{Results of CMM and CREML procedure for real data study}
\label{table:5}
\end{table}

In Table 5, estimates of all parameters are close between the CMM procedure and the CREML procedure. This is also because the data size is enough $(m=12, n=173)$. The estimates of $k_{nf}$ and $k_{pf}$ are consistent among the three procedures. The standard deviation of $\mu$ and $\lambda$ are smaller in the CREML procedure than in the CMM procedure, which implies the CREML procedure provides more accuracy. Moreover, there is a large positive correlation among $\mu$ and $\lambda$, which can be explained by the fact that the positive feedback process and negative feedback process in the first stage of tip growth process has an intrinsic connection since the strength of them both depend on the intensities of active ROP1 on the plasmic membrane.   

\section*{6. Discussion}
In this paper, we proposed an estimation procedure, CNLS for constrained nonlinear model and two estimation procedures, CMM and CREML for constrained nonlinear mixed model. This was initially motivated from an IDE based parameter estimation problem developed in tip growth process in developmental biology. However, they can also be used in any general constrained modeling problem. All the three procedures perform pretty well when the sample size is sufficiently large, whereas CREML outperforms CMM when the sample size is small. We used a simple strategy to incorporate the constraints into the objective function before applying simplex method to solve the constrained optimization problem in the estimation procedures, which works quite well.Other optimization methods such as Sequential Quadratic Programming can also be utilized. 

The methodology and theoretical result (Proposition 3) for the CNLS estimates are obtained by treating the differential equation parameter estimation problem as the standard nonlinear regression problem which usually has a closed-form objective function. In general, however, differential equation parameter estimation requires numerically solving the differential equation to evaluate the objective function, which produces a higher computational cost and additional numerical error. To deal with the local solution problem, the global optimization problem may need to be considered. Denote $h=max_{1\leq j\leq m-1}|X_{j+1}-X_j|$ as the maximum interval between samples. If there exists a $\gamma>0$ such that $h=O(n^{-\gamma})$ and the constrained area is bounded with the true parameters $\mu_0$ and $\lambda_0$ in the constrained area, then the estimators $\hat{\mu}_n,\hat{\lambda}_n$ will converge to $\mu_0$ and $\lambda_0$ almost surely, according to Theorem 3.1 of \cite{Xue:2010}.  This result accounts for the numerical error in solving differential equations.

The proposed CMM procedure is a standard “two-stage” method, which is not efficient. Although the proposed CREML method is better, the REML method for nonlinear mixed effects models is not easy to converge to the global solution when the parameter space is high. One solution to solve this problem is to use the result of CMM as starting value as we did in the paper.


\section*{Conflict of Interests Statement}
The authors have declared no conflict of interest.

\newpage

\appendix
\section{Appendix: Proofs}

\subsection{Proof of Lemma 1}
Based on the classical theory of the differential equation, there are potentially two solutions to the semilinear elliptical equation (3) including the null solution. Therefore, to prove Lemma 1, one only needs to show that there exists a non-null solution $\sigma_{0}$ to equation (3), and $\sigma_{0}>0$ on $[-c,c]$. 
 
\paragraph{}
The existence of a positive solution to the semilinear elliptic equation $-\partial_{x}^{2}u=f(u)$ is discussed in \cite{1982}. In our case, $f(u)=u^{\alpha}-u$. Therefore, $f(0)=0$, $f'(0)=-1<0$, and $f(u)$ is superlinear since $\frac{f(u)}{u} \rightarrow \infty$ as $u \rightarrow \infty$. By the Theorem 1.1 in \cite{1982}, there exists a positive function $\sigma_{0}$ in $C^{2}\left(\left[-c,c\right]\right)$ that satisfies equation (3). Furthermore, when $c= +\infty$, the existence and uniqueness of solution to the equation (3) can also be proved by the Theorem 1.1.3 in \cite{CaH98}. From \cite{Gidas1979Symmetry}, it is easy to see that $\sigma_{0}$ is a positive and even function which increases at $\left[-c,0\right]$ and decreases at $\left[0,c\right]$.

\subsection{Proof of Lemma 2}  		
Similar as the proof of Lemma 1, one only needs to show that there exists a non-null solution $R_{\lambda,\mu}(x)$ to equation (4), and $R_{\lambda,\mu}(x)>0$ on $[-L_0,L_0]$. 

Consider a family of functions $R_{\lambda,\mu}(x)=\lambda\sigma_{0}(\mu x)$ where $\lambda>0$, $\mu>0$, and $\sigma_{0}$ is the unique positive solution to equation (3) for $c=\mu L_0$. Then, 
\begin{align*}
\partial_{x}R_{\lambda,\mu}&=\lambda\mu\partial_{x}\sigma_{0}(\mu x)\\
\partial_{x}^{2}R_{\lambda,\mu}&=\lambda\mu^{2}\partial_{x}^{2}\sigma_{0}(\mu x)
\end{align*}
By equation (3), 
\begin{align*}
-\partial_{x}^{2}R_{\lambda,\mu}&=\lambda^{1-\alpha}\mu^{2}R_{\lambda,\mu}^{\alpha}-\mu^{2}R_{\lambda,\mu}.
\end{align*}

Therefore, $R_{\lambda,\mu}$ satisfies $-D\partial_{x}^{2}R_{\lambda,\mu}=-D\mu^{2}R_{\lambda,\mu}+D\lambda^{1-\alpha}\mu^{2}R_{\lambda,\mu}^{\alpha}$. Since $\mu, k_{nf}, k_{pf}, D>0$, we can take $\mu=\sqrt{\frac{k_{nf}}{D}}$ and $\lambda=\left(\frac{k_{pf}}{k_{nf}}\lambda'\right)^{\frac{1}{1-\alpha}}$, then $-D\partial_{x}^{2}R_{\lambda,\mu}=-k_{nf} R_{\lambda,\mu}+\lambda' k_{pf} R_{\lambda,\mu}^{\alpha}$. Therefore, $R_{\lambda,\mu}$ is the unique positive solution to equation (4).

\subsection{Proof of Sufficient Condition}
\textit{Proof.}
Since $R_{\lambda'}(x)=R_{\lambda,\mu}(x)$ is a solution to (\ref{eq:ParametrizedSemilinearElliptic}), $R_{\lambda'}(x)$ is also a solution to (\ref{eq:IDE}) if $\lambda'=\frac{k_{nf}}{k_{pf}}\lambda^{1-\alpha}  =  \left(1-\frac{1}{R_{tot}}\left\Vert R_{\lambda'}\right\Vert _{1}\right)$, where $\left\Vert R_{\lambda'}\right\Vert _{1} = \int_{-L_0}^{L_0}R_{\lambda,\mu}(x)dx = \lambda\sqrt{\frac{D}{k_{nf}}}\left\Vert \sigma_{0}\right\Vert _{1}$. Denote $g(\lambda)\doteq\frac{k_{nf}}{k_{pf}}-\lambda^{\alpha-1}+\frac{1}{R_{tot}}\lambda^{\alpha}\sqrt{\frac{D}{k_{nf}}}\left\Vert \sigma_{0}\right\Vert _{1}$, then $g'(\lambda)=\lambda^{\alpha-2}\left(-(\alpha-1)+\frac{\alpha}{R_{tot}}\sqrt{\frac{D}{k_{nf}}}\left\Vert \sigma_{0}\right\Vert _{1}\lambda\right)$.
The root $\lambda_{c}$ of $g'(\lambda)$ is $\lambda_{c}=\frac{\alpha-1}{\alpha}\sqrt{\frac{k_{nf}}{D}}\frac{R_{tot}}{\left\Vert \sigma_{0}\right\Vert _{1}}$, and $g(\lambda)$ is decreasing in $\left[0,\lambda_{c}\right]$ and increasing in $\left[\lambda_{c},+\infty\right]$. Notice that $g(0)=\frac{k_{nf}}{k_{pf}}$, $\lim_{+\infty}g=+\infty$, and 
\begin{align*}
g(\lambda_{c})=\frac{k_{nf}}{k_{pf}}-\frac{1}{\alpha}\left(\frac{\alpha-1}{\alpha}\sqrt{\frac{k_{nf}}{D}}\frac{R_{tot}}{\left\Vert \sigma_{0}\right\Vert _{1}}\right)^{\alpha-1}
\end{align*} 
 
\begin{enumerate}
  \item When $g(\lambda_c)>0$, $g(\lambda)>0$, no positive solution to (\ref{eq:IDE}) can be found from the family of solutions to (\ref{eq:ParametrizedSemilinearElliptic}).
  \item When $g(\lambda_c)=0$, $g(\lambda)>0$ for $\lambda\neq \lambda_c$, therefore one positive solution $R_{\lambda_c,\mu}(x)$ to (\ref{eq:IDE})  can be found from the family of solutions to (\ref{eq:ParametrizedSemilinearElliptic}).
  \item When $g(\lambda_c)<0$, there exist $\lambda_1\in [0, \lambda_c]$ and $\lambda_2\in [\lambda_c, \infty]$ such that $g(\lambda_1)=0$ and $g(\lambda_2)=0$, therefore two positive solutions $R_{\lambda_1,\mu}(x)$ and $R_{\lambda_2,\mu}(x)$ to (\ref{eq:IDE}) can be found from the family of solutions to (\ref{eq:ParametrizedSemilinearElliptic}).
\end{enumerate}

\subsection{Proof of Necessary Condition}

\textit{Proof.}
It is only necessary to show that for any positive solution $R$ of (\ref{eq:IDE}) on $\left[-L_0,L_0\right]$, there exist $\lambda,\mu>0$ such that $\sigma_0(x)=\frac{1}{\lambda}R(\frac{x}{\mu})$ is a solution to (\ref{eq:SemiEllipticEqu}) on $\left[-\frac{L_0}{\mu},\frac{L_0}{\mu}\right]$. Denote $\bar{\lambda}=\frac{1}{\lambda}$, $\bar{\mu}=\frac{1}{\mu}$, then $\frac{\partial \sigma_0(x)}{\partial x}=\bar{\lambda}\bar{\mu}\frac{\partial R(\bar{\mu}x)}{\partial(\bar{\mu}x)}$ and $\frac{\partial^2 \sigma_0(x)}{\partial x^2}=\bar{\lambda}\bar{\mu}^2\frac{\partial^2 R(\bar{\mu}x)}{\partial(\bar{\mu}x)^2}$. $\sigma_0(x)$ is a solution to (\ref{eq:SemiEllipticEqu}) on $\left[-L_0\bar{\mu},L_0\bar{\mu}\right]$ if and only if 
\begin{eqnarray*}
 & & -\frac{\partial^2\sigma_0(x)}{\partial x^2} = -\sigma_0(x)+\sigma_0^\alpha(x)\\
 & & -\bar{\lambda}\bar{\mu}^{2}R''(\bar{\mu}x) = -\bar{\lambda}R(\bar{\mu}x)+\bar{\lambda}^{\alpha}R^{\alpha}(\bar{\mu}x)\\
 & & -\bar{\mu}^{2}\frac{k_{nf}R(\bar{\mu}x)}{D} + \bar{\mu}^{2}\frac{k_{pf}R^{\alpha}(\bar{\mu}x)}{D}\left(1-\frac{\int_{-L_0}^{L_0}R(x)dx}{R_{tot}}\right) =  -R(\bar{\mu}x)+\bar{\lambda}^{\alpha-1}R^{\alpha}(\bar{\mu}x)
\end{eqnarray*}
when $\bar{\mu}=\sqrt{\frac{D}{k_{nf}}}$, $\bar{\lambda}$ can be obtained by solving the equality 
\[
\frac{k_{pf}}{k_{nf}}R^{\alpha}(\bar{\mu}x)\left(1-\frac{\int_{-L_0}^{L_0}R(x)dx}{R_{tot}}\right)=\bar{\lambda}^{\alpha-1}R^{\alpha}(\bar{\mu}x)
\]
for which $\bar{\lambda}=\left[\frac{k_{pf}}{k_{nf}}\left(1-\frac{\int_{-L_0}^{L_0}R(x)dx}{R_{tot}}\right)\right]^{\frac{1}{\alpha-1}}$. Hence, $\bar{\lambda}$ exists if and only if $\frac{\int_{-L_0}^{L_0}R(x)dx}{R_{tot}}<1$. Suppose $\frac{\int_{-L_0}^{L_0}R(x)dx}{R_{tot}}\geq 1$, then the right hand side of equation (\ref{eq:IDE}) is nonpositive and therefore the left hand side of equation (\ref{eq:IDE}) must be nonpositive. That is, $R''(x)>0$. Therefore, $R(x)$ must be a convex function. This is impossible because $R(x)$ is a positive function with $R(-L_0)=R(L_0)=0$. Therefore, $\frac{\int_{-L_0}^{L_0}R(x)dx}{R_{tot}}<1$ always holds for $R(x)>0$ and $\bar{\lambda}$ exists, which completes the proof.

\subsection{Proof of Remark 2}	
By Lemma 1, $\sigma_{0}$ is a positive and even function which increases at $\left[-c,0\right]$ and decreases at $\left[0,c\right]$. Therefore, $R_{\lambda,\mu}(x)=\lambda\sigma_0(\mu x)$ preserves the same properties. Moreover, in the proof of Lemma 1, the function $f(x)=x^\alpha-x$ is such that $f(1)=0$, and $f(x)>0$ for $x>1$. Therefore, by Theorem 3.1 of \cite{1982}, $\max_x\sigma_{0}(x)>1$ . As a result, $\max_x R_{\lambda,\mu}(x)=\lambda\times\max_x\sigma_{0}(\mu x)>\lambda$.

\subsection{Proof of Proposition 1}  
\begin{lemma}
for any $\mu>0$ and $\lambda>0$, the function $h(\mu,\lambda)=\lambda^{\alpha-1}-\lambda^{\alpha}\frac{\left\Vert \sigma_{0}\right\Vert _{1}}{\mu R_{tot}}-\frac{1}{\alpha}\left( \frac{\alpha-1}{\alpha} \frac{\mu R_{tot}}{\left\Vert \sigma_{0}\right\Vert _{1}} \right)^{\alpha-1}$ is always non-positive.
\end{lemma}

\textit{Proof.}
For any fixed $\mu>0$, $h$ is a function of $\lambda$ whose first-order derivative is 0 if and only if $\lambda \doteq \lambda_c =\frac{\alpha-1}{\alpha}\frac{\mu R_{tot}}{\left\Vert \sigma_{0}\right\Vert _{1}}$. Then, we have
\begin{align*}
 & h(\lambda_c) =\lambda_c^{\alpha-1}-\lambda_c^{\alpha}\frac{\left\Vert \sigma_{0}\right\Vert _{1}}{\mu R_{tot}}-\frac{1}{\alpha}\left( \frac{\alpha-1}{\alpha} \frac{\mu R_{tot}}{\left\Vert \sigma_{0}\right\Vert _{1}} \right)^{\alpha-1}=0 \\
 & h(0) =-\frac{1}{\alpha}\left( \frac{\alpha-1}{\alpha} \frac{\mu R_{tot}}{\left\Vert \sigma_{0}\right\Vert _{1}} \right)^{\alpha-1}<0\\
 & h(+\infty) =-\infty<0
\end{align*}
Notice that $h(\lambda)$ is a continuous function of $\lambda$, we can conclude that $h(\lambda)\leq 0$ based on the above three equations. Therefore, Lemma 3 holds.

When the constraints in model (7) are satisfied, $g(\lambda)$ has at least one solution. As a result, $\frac{k_{nf}}{k_{pf}}=\lambda^{\alpha-1}-\frac{1}{R_{tot}}\lambda^{\alpha}\frac{1}{\mu}\left\Vert \sigma_{0}\right\Vert _{1} > 0$ and $\mu>0$ and $\lambda>0$. Therefore, the constraints in model (8) hold. When the constraints in model (8) are satisfied, we can convert $\mu$ and $\lambda$ to $k_{nf}$ and $k_{pf}$ by solving $k_{nf}=D \mu^2$ and $\frac{k_{nf}}{k_{pf}}=\lambda^{\alpha-1}-\frac{1}{R_{tot}}\lambda^{\alpha}\frac{1}{\mu}\left\Vert \sigma_{0}\right\Vert _{1}$. The solution of $k_{nf}$ and $k_{pf}$ is such that $k_{nf}>0$, $k_{pf}>0$ and by Lemma 3, $\Lambda(k_{nf},k_{pf},D,R_{tot},\sigma_{0})=\frac{k_{nf}}{k_{pf}}-\frac{1}{\alpha}(\frac{\alpha-1}{\alpha}\sqrt{\frac{k_{nf}}{D}}\frac{R_{tot}}{\left\Vert \sigma_{0}\right\Vert _{1}})^{\alpha-1} = \lambda^{\alpha-1}-\lambda^{\alpha}\frac{\left\Vert \sigma_{0}\right\Vert _{1}}{R_{tot}\mu}-\frac{1}{\alpha}\left( \frac{\alpha-1}{\alpha} \frac{R_{tot}\mu}{\left\Vert \sigma_{0}\right\Vert _{1}} \right)^{\alpha-1} \leq 0$. Therefore, the constraints in model (7) hold.

\subsection{Proof of Proposition 2}  
\begin{lemma}
Let $A=(a_{ij})_{2\times 2}$ denote a symmetric two by two matrix. Suppose all the four elements of $A$ are bounded in $[-B,B]$ for some $B>0$, then $A \leq 2BI_2$. 
\end{lemma}

\textit{Proof.}
For any vector $\boldsymbol{x}=(x_1,x_2)^T$, $\boldsymbol{x}^TA\boldsymbol{x} = a_{11}x_1^2+2a_{12}x_1x_2+a_{22}x_2^2 \leq 2Bx_1^2+2Bx_2^2 =2B\boldsymbol{x}^T \boldsymbol{x}$. Therefore, $A \leq 2BI_2$ and Lemma 4 holds.

Denote $z=(z_\mu,z_\lambda)^T=n^{1/2}(\boldsymbol{\theta}-\boldsymbol{\theta_0})$. It can be easily seen that minimizing (\ref{Sec4leastsquare}) under the constraint (\ref{Sec4Constraint}) is equivalent to 
\begin{equation} \label{eq:28}
\begin{array}{c}
\min\limits_{z} \sum_{j=1}^n\{[\epsilon_j+R(x_j,\boldsymbol{\theta_0})-R(x_j,\boldsymbol{\theta_0}+n^{-1/2}z)]^2-\epsilon_j^2\}\\
\texttt{s.t.} g_1(\boldsymbol{\theta_0}+n^{-1/2}z)=-(\mu_0+n^{-1/2}z_\mu) R_{tot}+(\lambda_0+n^{-1/2}z_\lambda)\parallel\boldsymbol{\sigma_0}\parallel_1<0 \\
g_2(\boldsymbol{\theta_0}+n^{-1/2}z)=-(\mu_0+n^{-1/2}z_\mu)<0 \\
g_3(\boldsymbol{\theta_0}+n^{-1/2}z)=-(\lambda_0+n^{-1/2}z_\lambda)<0
\end{array}
\end{equation}
where $\epsilon_j$ are i.i.d with $N(0,\sigma^2)$. 

Assume the optimal solution of (\ref{eq:28}) exists and denote it by $\hat{z}_n$. Then $\hat{z}_n=n^{1/2}(\hat{\boldsymbol{\theta}}_n-\boldsymbol{\theta}_0)$. Therefore to prove proposition 2, we only need to prove $\hat{z}_n\xrightarrow{d}N(0,\sigma^2K^{-1})$, which can be achieved in the following two steps. First, we prove when $n\rightarrow\infty$ the limit problem of problem (\ref{eq:28}) is 
\begin{equation} \label{eq:27}
\texttt{min } z'Kz-2z'\xi
\end{equation} 
where $\xi\sim N(0,\sigma_\epsilon^2K)$. Then, we prove the solution to problem (\ref{eq:28}) converges in distribution to the solution to problem (\ref{eq:27}).

\subsection*{Step 1: Limit problem of (\ref{eq:28})}
Denote the objective function $F_n(\epsilon,z)=\sum_{j=1}^n\{[\epsilon_j+R(x_j,\boldsymbol{\theta_0})-R(x_j,\boldsymbol{\theta_0}+n^{-1/2}z)]^2-\epsilon_j^2\}$ and parameter space $S_n=\{z:g_1(\boldsymbol{\theta_0}+n^{-1/2}z)<0,g_2(\boldsymbol{\theta_0}+n^{-1/2}z)<0,g_3(\boldsymbol{\theta_0}+n^{-1/2}z)<0\}$. To formulate the limit problem of (\ref{eq:28}), we have the following results.

\textbf{Result 1:}  When $\sigma^2=1$, for each fixed $z\in \mathbb{R}^2$, $F_n(\epsilon,z)$ converges in distribution to $F(\xi,z)=z'Kz-2z'\xi$, where $\xi\sim N(0,K)$.

(i) As specified in Section 4, $\epsilon_1, \epsilon_2, \dots, \epsilon_n$ are $i.i.d.$ with $E(\epsilon_i)=0$ and $Var(\epsilon_i)=\sigma^2=1$.

(ii) $R(x_j;\boldsymbol{\theta})=\lambda\sigma_0(\mu x_j)$, $j=1,\dots,n$, are differentiable in $\boldsymbol{\theta}$ since $\sigma_0(\mu x_j)$ is differentiable in $\mu$. By Taylor expansion,
\begin{equation*}
R(x_j;\boldsymbol{\theta})=R(x_j;\boldsymbol{\theta}_0)+\nabla_{\boldsymbol{\theta}}R(x_j;\boldsymbol{\theta}_0)^T(\boldsymbol{\theta}-\boldsymbol{\theta}_0)+ \frac{1}{2}(\boldsymbol{\theta}-\boldsymbol{\theta}_0)^T \Delta_{\boldsymbol{\theta}}R(x_j;\boldsymbol{\theta}_0) (\boldsymbol{\theta}-\boldsymbol{\theta}_0)+o(\|\boldsymbol{\theta}-\boldsymbol{\theta}_0\|^2)  
\end{equation*}
where $\bigtriangledown_{\boldsymbol{\theta}} R(x_j;\boldsymbol{\theta_0})=\left(\begin{array}{c}
\lambda_0x_j\sigma_0'(\mu_0x_j) \\
\sigma_0(\mu_0x_j)
\end{array}\right)
$ and $\bigtriangleup_{\boldsymbol{\theta}} R(x_j;\boldsymbol{\theta_0})=
\left(\begin{array}{cc}
\lambda_0x_j^2\sigma_0''(\mu_0x_j) & x_j\sigma_0'(\mu_0x_j)\\
x_j\sigma_0'(\mu_0x_j) & 0
\end{array}\right)$. 

Let $r_j(\boldsymbol{\theta})=\{R(x_j;\boldsymbol{\theta})-R(x_j;\boldsymbol{\theta_0})-\bigtriangledown_{\boldsymbol{\theta}} R(x_j;\boldsymbol{\theta_0})^T(\boldsymbol{\theta}-\boldsymbol{\theta_0})\}/\parallel\boldsymbol{\theta}-\boldsymbol{\theta_0}\parallel^2$. Since $R(x_j;\boldsymbol{\theta})$, $\bigtriangledown_{\boldsymbol{\theta}} R(x_j;\boldsymbol{\theta_0})^T(\boldsymbol{\theta}-\boldsymbol{\theta_0})$,$\parallel\boldsymbol{\theta}-\boldsymbol{\theta_0}\parallel^2$ are continuous on $\boldsymbol{\theta}$, $r_j(\boldsymbol{\theta})$ is a continuous function on $\boldsymbol{\theta}$.

It's obvious that there exists $B>0$ such that all elements in $\bigtriangleup_{\boldsymbol{\theta}} R(x_j,\boldsymbol{\theta_0})$ are bounded by $[-B,B]$. Therefore, from Lemma 4, we have 
\begin{eqnarray*}
 |r_j(\boldsymbol{\theta})|\parallel\boldsymbol{\theta}-\boldsymbol{\theta}_0\parallel^2 &=&|\frac{1}{2}(\boldsymbol{\theta}-\boldsymbol{\theta}_0)^T \Delta_{\boldsymbol{\theta}}R(x_j;\boldsymbol{\theta}_0) (\boldsymbol{\theta}-\boldsymbol{\theta}_0)+o(\|\boldsymbol{\theta}-\boldsymbol{\theta}_0\|^2)|\\
& \leq & \frac{1}{2}(\boldsymbol{\theta}-\boldsymbol{\theta}_0)^T 2BI_2 (\boldsymbol{\theta}-\boldsymbol{\theta}_0)\\
&=&B\|\boldsymbol{\theta}-\boldsymbol{\theta}_0\|^2  
\end{eqnarray*} 

Therefore, $|r_j(\boldsymbol{\theta})| \leq B$ and $\lim_{n \to \infty} \frac{1}{n}\sum_{j=1}^{n} r_j^2(\boldsymbol{\theta}) \leq B^2 < \infty$ holds in the whole parameter space. 

(iii) Since $\bigtriangledown_{\boldsymbol{\theta}} R(x_j,\boldsymbol{\theta}_0)\bigtriangledown_{\boldsymbol{\theta}} R(x_j,\boldsymbol{\theta}_0)'=\left(\begin{array}{cc}
\lambda_0^2x_j^2\sigma_0'(\mu_0x_j)^2 & \lambda_0x_j\sigma_0(\mu_0x_j)\sigma_0'(\mu_0x_j) \\
\lambda_0x_j\sigma_0(\mu_0x_j)\sigma_0'(\mu_0x_j) & \sigma_0(\mu_0x_j)^2
\end{array}\right)$, all the elements in $\bigtriangledown_{\boldsymbol{\theta}} R(x_j,\boldsymbol{\theta}_0)\bigtriangledown_{\boldsymbol{\theta}} R(x_j,\boldsymbol{\theta}_0)'$ are bounded. By Kolmogorov's Strong Law of Large Numbers (SLLN), we have
\begin{equation*}
\dfrac{1}{n}\sum_{j=1}^n\bigtriangledown_{\boldsymbol{\theta}} R(x_j,\boldsymbol{\theta}_0)\bigtriangledown_{\boldsymbol{\theta}} R(x_j,\boldsymbol{\theta}_0)'\xrightarrow{a.s.}K
\end{equation*}
where
\begin{equation*}
K=\left(\begin{array}{cc}
E_X[\lambda_0^2X^2\sigma_0'(\mu_0X)^2] & E_X[\lambda_0X\sigma_0(\mu_0X)\sigma_0'(\mu_0X)] \\
E_X[\lambda_0X\sigma_0(\mu_0X)\sigma_0'(\mu_0X)] & E_X[\sigma_0(\mu_0X)^2]
\end{array}\right)
\end{equation*}

By Cauchy-Schwarz inequality we have
\begin{equation*}
det(K)=E_X[\lambda_0^2X^2\sigma_0'(\mu_0X)^2]
E_X[\sigma_0(\mu_0X)^2]-
(E_X[\lambda_0X\sigma_0(\mu_0X)\sigma_0'(\mu_0X)])^2\ge 0
\end{equation*}

However, if equality holds, it implies that $\lambda_0X\sigma_0'(\mu_0X)$ and $\sigma_0(\mu_0X)$ are linearly dependent, i.e., there exists a non-zero scalar $a\in R$ such that $\lambda_0 X \sigma_0'(\mu_0 X)=a\sigma_0(\mu_0 X)$ holds everywhere since $\sigma_0(\mu_0 X)$ and $\sigma_0'(\mu_0 X)$ are both continuously differentiable.  As a result, $\sigma_0(\mu_0 X)$ is a solution to the linear ODE $\lambda_0 u'(X)-a u(X)=0$. However, the solution is $u(X)=Ce^{\frac{a}{\lambda_0}x}$ which can not satisfy the boundary condition required for $\sigma_0(X)$. Therefore $det(K)>0$. Since $trace(K)>0$, both eigenvalues of $K$ are positive. So $K$ is positive definite.

Therefore, $\lim_{n\to\infty}\dfrac{1}{n}\sum_{j=1}^n\bigtriangledown_{\boldsymbol{\theta}} R(x_j,\boldsymbol{\theta}_0)\bigtriangledown_{\boldsymbol{\theta}} R(x_j,\boldsymbol{\theta}_0)'=K$ exists and is positive definite.

From Theorem 1 of \cite{citeulike:1392995}, we have $F_n(\epsilon,z)$ converges in distribution to $F(\xi,z)=z'Kz-2z'\xi$.

\paragraph{Result 2:} It is obvious that $g_i(\boldsymbol{\theta})$ are continuously differentiable and there exists no equality constraints.  Also because\\
$\left\lbrace\begin{array}{c}
g_1(\boldsymbol{\theta}_0)=\mu_0 R_{tot}-\lambda_0\parallel\sigma_0\parallel_1\neq0 \\
g_2(\boldsymbol{\theta}_0)=\mu_0\neq0 \\
g_3(\boldsymbol{\theta}_0)=\lambda_0\neq0
\end{array}\right.$. \\
$I$ is an empty set. Therefore, by theorem 2 of \cite{citeulike:1392995}, we have parameter space $S_n$ converges in Kuratowski's sense to $S$ which is the parameter space of \ref{eq:27}. Combining part 1 and part 2, the limit problem of (\ref{eq:28}) is minimizing $z'Kz-2z'\xi$ without constraint.

\subsection*{Step 2 Convergence of solution to (\ref{eq:28})}

According to theorem 3-6 of \cite{citeulike:1392995}, the solution to limit problem \ref{eq:27} should be unique at $B(M)={z:\parallel z\parallel<M}$ for any large $M$, so that the solution to (\ref{eq:28}) converges in distribution to the solution to (\ref{eq:27}),

Since limit problem (\ref{eq:27}) is minimizing $z'Kz-2z'\xi$ without constraint, there is a unique solution $z=K^{-1}\xi$ at $B(M)={z:\parallel z\parallel<M}$ for any $M>\parallel K^{-1}\xi\parallel$. Therefore, by theorem 3-6 of \cite{citeulike:1392995}, $\hat{z}_n$ of problem (\ref{eq:28}) converges in distribution to $z=K^{-1}\xi\sim N(0,\sigma^2K^{-1})$, i.e. $\hat{z}_n=\sqrt{n}(\hat{\theta}_n-\theta_0)\xrightarrow{d}N(0,\sigma^2K^{-1})$. This completes the proof.

For any $\sigma^2>0$, based on Theorem 4 and 5 of Jennrich \cite{citeulike:12197486}, 
\begin{equation*}
n^{-\frac{1}{2}}\sum_{i=1}^n \nabla_{\boldsymbol{\theta}}R(X_i;\boldsymbol{\theta}_0) \boldsymbol{\epsilon}_i \stackrel{d}{\rightarrow} N(0, \sigma^2 K) 
\end{equation*}

As a result, $F_n(\boldsymbol{\epsilon}_i,\boldsymbol{z})$ will converge in distribution to $\boldsymbol{z}'K\boldsymbol{z}-2\boldsymbol{z}'\boldsymbol{\xi}$ where $\boldsymbol{\xi}$ is a random vector following $N(0,\sigma^2 K)$. In fact, based on Theorem 1-5 of Jennrich \cite{citeulike:12197486}, Theorem 1-6 of \cite{citeulike:1392995} still hold even if $\sigma^2$ is unknown.

\subsection{Proof of Proposition 3}

In this section, we want to prove the consistency of $\hat{\sigma}_{n}^2=\dfrac{1}{n}\sum_{j=1}^n(y_j-R(x_j,\hat{\boldsymbol{\theta}}_n))^2$, i.e. $\hat{\sigma}_{n}^2\xrightarrow{p}\sigma^2$.

\begin{equation} \label{eq:31}
\begin{split}
\hat{\sigma}_{n}^2 & =\dfrac{1}{n}\sum_{j=1}^n(y_j-R(x_j,\hat{\boldsymbol{\theta}}_n))^2=\dfrac{1}{n}\sum_{j=1}^n(R(x_j,\boldsymbol{\theta}_0)+\epsilon_j-R(x_j,\hat{\boldsymbol{\theta}}_n))^2\\
 & =\dfrac{1}{n}\sum_{j=1}^n(R(x_j,\boldsymbol{\theta}_0)-R(x_j,\hat{\boldsymbol{\theta}}_n))^2+\dfrac{2}{n}\sum_{j=1}^n(R(x_j,\boldsymbol{\theta}_0)-R(x_j,\hat{\boldsymbol{\theta}}_n))\epsilon_j+\dfrac{1}{n}\sum_{j=1}^n\epsilon_j^2
\end{split}
\end{equation}

First, we prove that $\dfrac{1}{n}\sum_{j=1}^n(R(x_j,\boldsymbol{\theta}_0)-R(x_j,\hat{\boldsymbol{\theta}}_n))^2\xrightarrow{p}0$. From proof of Proposition 2, we have $\dfrac{1}{n}\sum_{j=1}^n\bigtriangledown_{\boldsymbol{\theta}} R(x_j,\boldsymbol{\theta}_0)\bigtriangledown_{\boldsymbol{\theta}} R(x_j,\boldsymbol{\theta}_0)^T\xrightarrow{a.s.}K$ and $K\leq 2BI$. Since $\sqrt{n}(\hat{\boldsymbol{\theta}}_n-\boldsymbol{\theta}_0)\xrightarrow{d}N(0,\sigma^2K^{-1})$, we have $\hat{\boldsymbol{\theta}}_n\xrightarrow{L_2}\boldsymbol{\theta}_0$. Therefore, we have
\begin{equation} \label{eq:32}
\begin{split}
&\dfrac{1}{n}\sum_{j=1}^n(R(x_j,\boldsymbol{\theta}_0)-R(x_j,\hat{\boldsymbol{\theta}}_n))^2\\
&=(\hat{\boldsymbol{\theta}}_n-\boldsymbol{\theta}_0)^T[\dfrac{1}{n}\sum_{j=1}^n\bigtriangledown_{\boldsymbol{\theta}} R(x_j,\boldsymbol{\theta}_0)\bigtriangledown_{\boldsymbol{\theta}} R(x_j,\boldsymbol{\theta}_0)^T](\hat{\boldsymbol{\theta}}_n-\boldsymbol{\theta}_0)+o(\parallel\hat{\boldsymbol{\theta}}_n-\boldsymbol{\theta}\parallel^2)\\
&\xrightarrow{p}0
\end{split}
\end{equation}

Since $\bigtriangledown_{\boldsymbol{\theta}} R(x_j,\boldsymbol{\theta}_0)=\left(\begin{array}{c}
\lambda_0x_j\sigma_0'(\mu_0x_j) \\
\sigma_0(\mu_0x_j)
\end{array}\right)$
which is continuous and bounded, by theorem 4 of \cite{citeulike:12197486}, we have $\dfrac{1}{n}\sum_{j=1}^n\bigtriangledown_{\boldsymbol{\theta}} R(x_j,\boldsymbol{\theta}_0)\epsilon_j\xrightarrow{p}0$. Therefore, we have

\begin{equation} \label{eq:33}
\dfrac{1}{n}\sum_{j=1}^n(R(x_j,\boldsymbol{\theta}_0)-R(x_j,\hat{\boldsymbol{\theta}}_n))\epsilon_j=(\hat{\boldsymbol{\theta}}_n-\boldsymbol{\theta}_0)^T\{\dfrac{1}{n}\sum_{j=1}^n\bigtriangledown_{\boldsymbol{\theta}} R(x_j,\boldsymbol{\theta}_0)\epsilon_j\}+o(\parallel\hat{\boldsymbol{\theta}}_n-\boldsymbol{\theta}_0\parallel)\xrightarrow{p}0
\end{equation}

By SLLN, $\dfrac{1}{n}\sum_{j=1}^n\epsilon_j^2\xrightarrow{a.s.}\sigma^2$. From equation (\ref{eq:31}), (\ref{eq:32}) and (\ref{eq:33}), we have $\hat{\sigma}_{n}^2\xrightarrow{p}\sigma^2$.

\subsection{Proof of Proposition 4}  
Since $\hat{\boldsymbol{\theta}_i}=(\hat{\mu}_i,\hat{\lambda}_i)^T$ is obtained by CNLS for each single pollen tube, from Proposition 3 we have 
\begin{equation*}
\sqrt{n}(\hat{\boldsymbol{\theta}}_i-\boldsymbol{\theta}_i)\stackrel{d}{\rightarrow} MVN \left( \mathbf{0}, \sigma^2 (K_i)^{-1} \right ) 
\end{equation*}  
for each given $\boldsymbol{\theta}_i$, where $K_i=E_{X}[\nabla_{\boldsymbol{\theta}_i}R(X;\boldsymbol{\theta}_i)\nabla_{\boldsymbol{\theta}_i}R(X;\boldsymbol{\theta}_i)^T]$. Since $\boldsymbol{\theta}_i\sim MVN(\boldsymbol{\theta},\Sigma)$, the unconditional asymptotic mean and variance of $\hat{\boldsymbol{\theta}}_i$ are
\begin{align*}
E(\hat{\boldsymbol{\theta}}_i) & = E_{\boldsymbol{\theta}}[E_\epsilon (\hat{\boldsymbol{\theta}}_i|\boldsymbol{\theta}_i)]\rightarrow E_{\boldsymbol{\theta}} (\boldsymbol{\theta}_i)=\boldsymbol{\theta} \\
Var(\hat{\boldsymbol{\theta}}_i) & = Var_{\boldsymbol{\theta}}[E_\epsilon (\hat{\boldsymbol{\theta}}_i|\boldsymbol{\theta}_i)] + E_{\boldsymbol{\theta}}[Var_\epsilon (\hat{\boldsymbol{\theta}}_i|\boldsymbol{\theta}_i)] \rightarrow Var_{\boldsymbol{\theta}}[\boldsymbol{\theta}_i] + E_{\boldsymbol{\theta}}[(n K_i)^{-1}\sigma^2 ]\\
                          & = \Sigma + \sigma^2 E_{\boldsymbol{\theta}}[(n K_i)^{-1}] \doteq \tilde{\Sigma}
\end{align*}
Therefore, $\{\hat{\boldsymbol{\theta}}_i : i=1,\cdots,m\}$ are $i.i.d.$ with common asymptotic mean and variance. Since $\hat{\boldsymbol{\theta}}=m^{-1}\sum\hat{\boldsymbol{\theta}}_i$, from SLLN and CLT, we have

\begin{equation*}
\begin{split}
\dfrac{1}{m-1}\sum_{i=1}^m(\hat{\boldsymbol{\theta}}_i-\hat{\boldsymbol{\theta}})(\hat{\boldsymbol{\theta}}_i-\hat{\boldsymbol{\theta}})^T\xrightarrow{p}\tilde{\Sigma}\\
\hat{\boldsymbol{\theta}} \stackrel{p}{\rightarrow} \boldsymbol{\theta}\\
\sqrt{m}\tilde{\Sigma}^{-\frac{1}{2}}(\hat{\boldsymbol{\theta}}- \boldsymbol{\theta}) \stackrel{d}{\rightarrow} \boldsymbol{Z}
\end{split}
\end{equation*}
with $\boldsymbol{Z} \sim N(0,I_2)$. 

Furthermore, we have
\begin{align*}
 & E(T_i^{-1}) = E_{\boldsymbol{\theta}}[E_\epsilon (T_i^{-1}|\boldsymbol{\theta}_i)] = E_{\boldsymbol{\theta}}[E_\epsilon (  \left( \left[\frac{\partial \boldsymbol{R}_i}{\partial \boldsymbol{\theta}_i^T}\right]^T\left[ \frac{\partial \boldsymbol{R}_i}{\partial \boldsymbol{\theta}_i^T} \right] |_{\boldsymbol{\theta}_i=\hat{\boldsymbol{\theta}}_i} \right)^{-1}|\boldsymbol{\theta}_i)] \\
                                  & \stackrel{p}{\rightarrow} E_{\boldsymbol{\theta}} \left[ \left( \left[\frac{\partial \boldsymbol{R}_i}{\partial \boldsymbol{\theta}_i^T}\right]^T\left[ \frac{\partial \boldsymbol{R}_i}{\partial \boldsymbol{\theta}_i^T} \right] \right)^{-1} \right] = E_{\boldsymbol{\theta}} \left[ \left(\sum_{j=1}^{n}\left[ \frac{\partial R(X_{ij};\boldsymbol{\theta}_i)}{\partial \boldsymbol{\theta}_i^T}\right]\left[\frac{\partial R(X_{ij};\boldsymbol{\theta}_i )}{\partial \boldsymbol{\theta}_i^T}\right]^T \right)^{-1} \right]\\
                                  & \stackrel{p}{\rightarrow} E_{\boldsymbol{\theta}} [ \left( n E_X [\nabla_{\boldsymbol{\theta}_i}R(X;\boldsymbol{\theta}_i)\nabla_{\boldsymbol{\theta}_i}R(X;\boldsymbol{\theta}_i)^T] \right)^{-1} ]= E_{\boldsymbol{\theta}} [(n K_i)^{-1}]
\end{align*}

The first ``$\stackrel{p}{\rightarrow}$'' in the above equation holds since $\hat{\boldsymbol{\theta}}_i\stackrel{p}{\rightarrow}\boldsymbol{\theta}_i$. The second ``$\stackrel{p}{\rightarrow}$'' holds by SLLN of $X$. Therefore, $\{T_i^{-1}: i=1,\cdots,m \}$ are $i.i.d.$ with the same asymptotic mean, and so by SLLN we have that $\frac{1}{m}\sum_{i=1}^m T_i^{-1} \stackrel{p}{\rightarrow} E_{\boldsymbol{\theta}}[(n K_i)^{-1}]$. In addition, it's assumed that $\hat{\sigma}^2 \stackrel{p}{\rightarrow} \sigma^2$. Therefore, by \textit{Slutsky's} Theorem, $\hat{\Sigma} \stackrel{p}{\rightarrow} \tilde{\Sigma}- \sigma^2 E_{\boldsymbol{\theta}}[(n K_i)^{-1}]=\Sigma$.

\paragraph{}
Based on the asymptotical result of $\hat{\Sigma}$, we know that $\hat{\tilde{\Sigma}}=\hat{\Sigma} + \hat{\sigma}^2 E_{\boldsymbol{\theta}}[(nK_i)^{-1}] \stackrel{p}{\rightarrow} \tilde{\Sigma}$. We also have proved that $\hat{\boldsymbol{\theta}}$, $\sqrt{m}(\hat{\boldsymbol{\theta}}- \boldsymbol{\theta}) \stackrel{d}{\rightarrow} \tilde{\Sigma}^{\frac{1}{2}}\boldsymbol{Z}$. Therefore, by \textit{Slutsky's} Theorem we have $\sqrt{m}\hat{\tilde{\Sigma}}^{-\frac{1}{2}}(\hat{\boldsymbol{\theta}}- \boldsymbol{\theta}) \stackrel{d}{\rightarrow} \boldsymbol{Z}$. This completes the proof of Proposition 4.

\newpage

\bibliographystyle{biom}
\bibliography{bibfile2b}

\end{document}